\documentclass[a4paper,fleqn]{cas-dc}

\usepackage{graphicx}
\usepackage{epstopdf}
\usepackage{subcaption}
\usepackage{etoolbox}
\usepackage{xspace}
\usepackage{tabularx}
\usepackage{booktabs}
\usepackage{multirow}
\usepackage{rotating}
\usepackage{ragged2e}
\usepackage{array}
\usepackage[table,x11names,svgnames,dvipsnames]{xcolor}
\usepackage{tikz}
\usetikzlibrary{plotmarks}
\usepackage{pgfplots}
\pgfplotsset{compat=1.18}
\usepackage{algorithm}
\usepackage{algpseudocode}
\usepackage{listings}
\usepackage{caption}
\usepackage[section]{placeins} 
\usepackage[numbers,sort&compress]{natbib}

\newcolumntype{Y}{>{\RaggedRight\arraybackslash}X}
\renewcommand{\arraystretch}{1.02}
\lstdefinestyle{bashstyle}{
  language=bash,
  basicstyle=\ttfamily\footnotesize,
  columns=fullflexible,
  breaklines=true,
  keepspaces=true,
  showstringspaces=false,
  frame=single,
  numbersep=8pt,
  xleftmargin=0pt
}

\lstdefinestyle{terminal}{
  basicstyle=\ttfamily\footnotesize,
  columns=fullflexible,
  breaklines=true,
  keepspaces=true,
  showstringspaces=false,
  frame=single
}

\providecommand{\ie}{i.e.\xspace}
\providecommand{\eg}{e.g.\xspace}
\providecommand{\etal}{\textit{et al.\@}\xspace}

\def\tsc#1{\csdef{#1}{\textsc{\lowercase{#1}}\xspace}}
\tsc{WGM}
\tsc{QE}
\tsc{EP}
\tsc{PMS}
\tsc{BEC}
\tsc{DE}

\begin{document}
\let\WriteBookmarks\relax

\renewcommand{\topfraction}{0.95}
\renewcommand{\dbltopfraction}{0.95}
\renewcommand{\textfraction}{0.05}
\renewcommand{\floatpagefraction}{0.80}
\renewcommand{\dblfloatpagefraction}{0.80}
\setcounter{topnumber}{5}
\setcounter{dbltopnumber}{3}
\setcounter{totalnumber}{7}

\shorttitle{MoLIFE}

\shortauthors{S.L. Sanna et~al.}

\title [mode = title]{MoLIFE: Methodology, Technologies, and Challenges for \\ Mobile Live Intelligent Forensics Examination}                      

\author[1]{Silvia Lucia Sanna}[type=editor,
                        auid=000,bioid=1]

\cormark[1]
\cortext[1]{Corresponding author}


\ead{silvial.sanna@unica.it}
\credit{Conceptualization, Investigation, Methodology, Validation, Visualization, Software, Formal Analysis, Data Curation, Resources, Writing}

\author[3]{Cristina Alcaraz}[type=editor,
                        auid=000,bioid=2]

\ead{alcaraz@uma.es}


\credit{Conceptualization, Methodology, Supervision, Writing (review), Project Administration, Validation, Visualization}

\author[1]{Alessandro Sanna}[type=editor,
                        auid=000,bioid=3]

\ead{alessandro.sanna96@unica.it}


\credit{Conceptualization, Methodology, Validation, Visualization, Formal Analysis, Writing (review)}

\author[1,2]{Giorgio Giacinto}[type=editor,
                        auid=000,bioid=4]

\ead{giorgio.giacinto@unica.it}


\credit{Supervision, Conceptualization, Funding, Project Administration, Methodology, Resources, Validation, Visualization, Writing (review)}

\author[3]{Javier Lopez}[type=editor,
                        auid=000,bioid=5]

\ead{javierlopez@uma.es}


\credit{Supervision, Writing (review), Validation, Visualization, Project Administration}


\affiliation[1]{organization={University of Cagliari, Dept of Electrical and Electronic Engineering},
    addressline={Via Marengo, 2}, 
    city={Cagliari},
    postcode={09123}, 
    country={Italy}}

\affiliation[2]{organization={Interuniversity National Consortium for Informatics, CINI},
    addressline={Via Ariosto, 25}, 
    city={Rome},
    postcode={00185}, 
    country={Italy}}

\affiliation[3]{organization={University of Malaga, NICS Lab},
    addressline={Campus de Teatinos s/n}, 
    city={Malaga},
    postcode={29071}, 
    country={Spain}}

\begin{abstract}
Nowadays, mobile forensics is less explored in Digital Forensics case analysis due to the increase in data protection mechanisms implemented by tech companies (\ie Google for Android and Apple for iOS). For example, the physical acquisition or analysis of specific directories under super-user protection would corrupt the evidence; access to such data is protected, and bypassing this protection requires either privilege escalation or custom ROM installation, leading to the modification of the device state. At the same time, the demand for mobile technologies and their respective communication systems is increasing exponentially, exposing numerous security threats and risks. For that reason, this paper presents a Mobile Live Intelligent Forensics Examination (MoLIFE), a novel Digital Forensics (DF) methodology for data acquisition and analysis of mobile devices. The proposed methodology is based on NIST SP800-101 for the DF process. MoLIFE can be integrated with new and emerging technologies by exploiting their power (\eg AI, blockchain, quantum computing). MoLIFE can also be used to prevent cyber threats and incidents, as well as DF post-mortem analysis, offering examples of applying the MoLIFE methodology and good practices for the future. To prove the technical feasibility of the methodology, a small case study on Android devices data acquisition via the mDT will be presented. As the methodology is based on new and emerging technologies, it depends on their limitations that would be overcome in a few years.
\end{abstract}




\maketitle

\section{Introduction}
\label{sec:intro}
Mobile Forensics is the science of retrieving data from mobile devices \cite{Alatawi20_ICCIT}. It includes a set of techniques to extract and analyze data stored in a mobile device when a specific crime is committed. Such analysis can be conducted only if requested and authorized by a judge. Nowadays, mobile devices are part of our everyday life, as well as tools for a wide variety of business-related tasks. In fact, they are widely employed for personal necessities or working activities (\eg industries, critical infrastructure monitoring, healthcare data collection and monitoring, surveillance, etc.). So far, there are two main Operating Systems (OS) for mobile devices: Android, developed by Google and installed by many vendors (\eg Samsung, Oppo, Xiaomi), and iOS, developed by Apple and installed in iPhones\footnote{\url{https://gs.statcounter.com/os-market-share/mobile/worldwide}}. The main differences between such devices do not rely only on the OS used but also on the vendor \cite{Sahani17_JCA}. In fact, different vendors can introduce distinct protection mechanisms, proprietary Software Applications (SwA), and custom hardware platforms. Reports show that there are $7.2$ billion of mobile devices with millions of people connected worldwide\cite{howarth2025how}, expected to grow with an investment of USD $6.79$ Billion for $2029$ \cite{databridge2022global}. The number of cyber attacks on mobile devices has thus increased rapidly in recent years \cite{Hwaitat21}.

Regarding cyber attacks, a Digital Forensics (DF) analysis is necessary for the incident response part, especially when dealing with a mobile device used in a business-related context such as an industry or company, particularly if related to critical infrastructures. A mobile device is often forensically analyzed even when a civil or penal crime is committed. Although the mobile device is not strictly related to the crime (\ie it is not the first actor of the crime as in the case of a cybercrime), it is analyzed in the civil/penal trial because it can contain important information related to the case. For example, if a pseudo-pedopornography crime is committed, a mobile device could include pictures, chats, and website history; and for that reason, it must be analyzed. Due to the different vendors and proprietary systems, the analysis is not universal and relies strictly on them. Moreover, super-user permission (\ie root, admin) is sometimes required to access specific data stored on the device. The acquisition of the super-user privilege strictly depends on the OS type and version, vendor type, and hardware specifics. In fact, most of the time, the mobile device is not analyzed because the acquisition of such permission compromises the entire device \cite{Fakiha24_JISIS}. However, some DF tools such as {\tt Ufed}\footnote{\url{https://cellebrite.com/en/ufed}}, (\ie an Israeli digital intelligence firm) and {\tt Oxygen}\footnote{\url{https://www.oxygenforensics.com}} (\ie a US-based investigative technology company) can dump a wide range of information even if the device is non-rooted, by using specific exploits to grant privileged access. This leads to a lack of trustworthiness derived from the black-box techniques not reliable in trials, as they do not transpire the integrity of the evidence. Due to the new adopted security patches and data protection mechanisms regarding privacy and countermeasures for cyberattacks and vulnerabilities, such tools are becoming less effective in recent devices. 

As described by the National Institute of Standards and Technology (NIST) in the standard SP800-101~\cite{Ayers14_NIST}, one of the main principles of DF is the reliability of the data, including preservation of the state of the evidence (\ie not corrupting the device and its data), the reproducibility and the availability of the data \cite{Cuomo22_Sensors}. Not acquiring super-user permissions properly sometimes leads to a data factory reset with a consequent data deletion, and such DF principle is violated. Therefore, the evidence cannot be analyzed and admitted in the trial. However, if the device is not analyzed, the case is partially solved from the legal perspective, as the device can contain essential data. Specifically, mobile devices are rarely physically analyzed, \ie a bit-by-bit copy of the evidence at the seizure time to be fully replicated and analyzed in every aspect. Sometimes, if the owner is willing to cooperate (\ie a company in the case of a cyber incident or an innocent defendant), the secret message protecting the device is given, and a live forensics analysis of specific parts without super-user permission is made. In all other cases, modern forensics techniques must be developed explicitly for mobile investigation, such as bypassing security features and exploiting system vulnerabilities to read data \cite{Fukami21_FSIDI}. Potentially, such methodology can lead to data integrity violations and should be supported by legislative frameworks. 

\begin{figure}[pos=t]
    \centering
    \includegraphics[width=\linewidth]{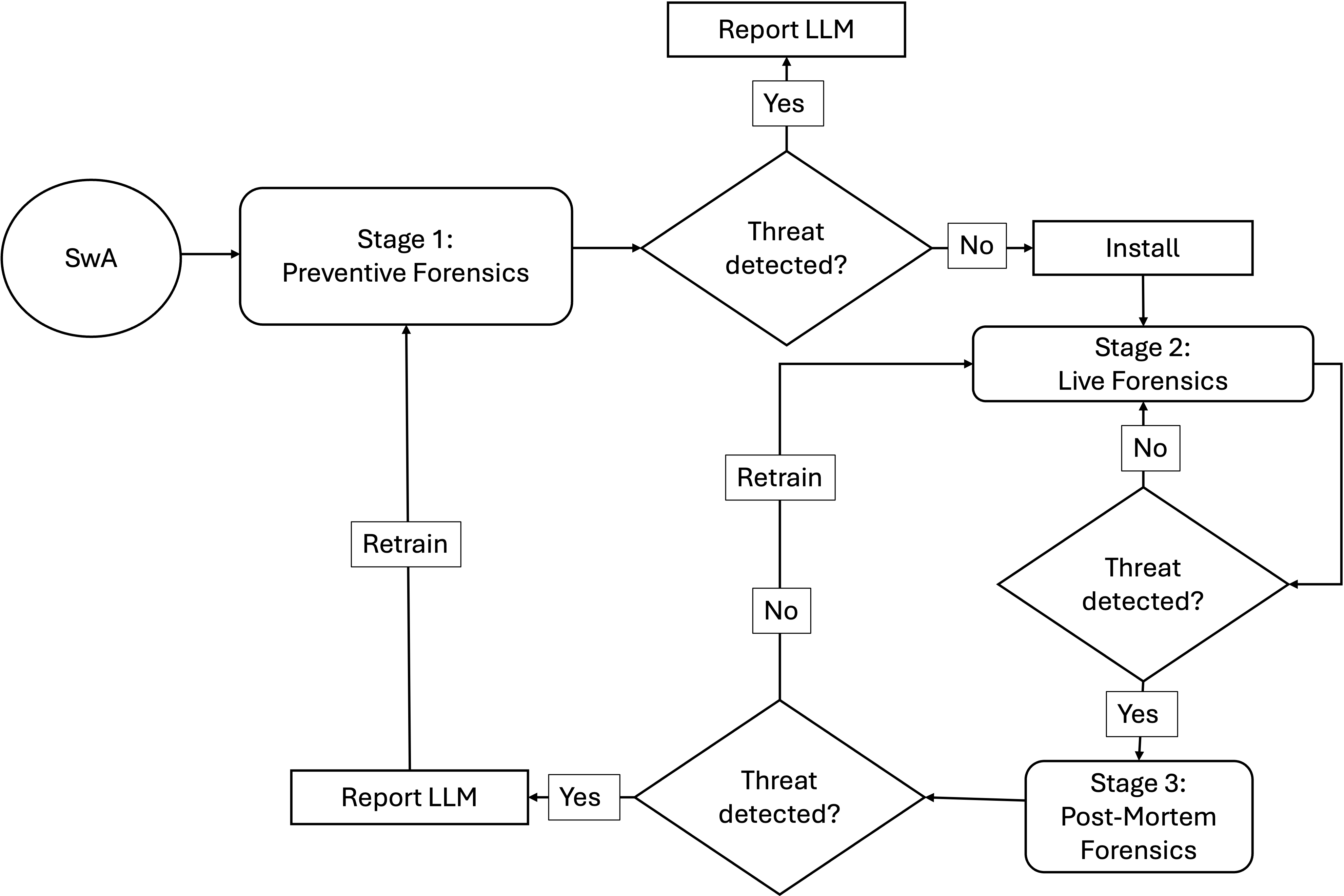}
    \caption{MoLIFE flow diagram including all three mDT stages. Given a mobile SwA, first check if it is a known popular threat (preventive analysis) and, in case, report it. If the SwA is secure, it monitors its execution (real-time monitoring) and, once it detects the threat, reports the incident and finds the causes (post-mortem analysis)}
    \label{fig:flowdiagram}
\end{figure}

Commercial forensic tools (\eg {\tt Magnet AXIOM}, {\tt Cellebrite}, {\tt Oxygen}) integrate acquisition, analysis, and reporting, and can often extract data from unrooted devices using backups, system APIs, or temporary agents, escalating to exploits or hardware methods only when needed. Additionally, such tools sometimes are not reliable from the juridical perspective because of their black-box implementation for rooting privileges. Academic research has explored non-root/jailbreak acquisitions (logical, backup-based, or SwA-specific), which preserve forensic integrity but remain incomplete, as they cannot access deleted, encrypted, or system-level data. Overall, non-privileged methods are safer for evidentiary purposes but must be complemented with privileged techniques for full data acquisition. Because extraction methods vary by device model, OS build and platform protections (\eg Secure Enclave/Activation Lock on iOS), any privileged access must be legally authorized, exhaustively documented (tools, versions, commands, hashes) and, where possible, independently validated to preserve evidentiary integrity. Native backup methods are non-invasive but incomplete and inconsistent, while approaches that bypass encryption via vulnerability exploitation (\eg Fukami \etal~\cite{Fukami21_FSIDI}) can extend access but rely on OS-specific flaws, \ie non-generalizable and non-scalable. To overcome these issues, this paper proposes a new methodology for Mobile Live Forensics Examination (MoLIFE), based on the NIST DF main principles, to analyze mobile devices even without acquiring super-user permissions on the real device. The MoLIFE methodology, Figure~\ref{fig:flowdiagram}, can be applied in industries, companies, and critical infrastructures to monitor and prevent cyber incidents. Specifically, the MoLIFE methodology can support continuous monitoring, ensuring complete certainty on what happened. The methodology can be applied in some private investigations when the device owner collaborates for a complete acquisition of the current state of the smartphone, by reading data under super-user privileges without damaging or invalidating the evidence. 

More specifically, the main pillar of MoLIFE methodology is a system of Digital Twins (DT) \cite{Abdelrahman25_BE}, by applying the concept to mobile devices, introducing the concept of mobile Digital Twin (mDT). MoLIFE can be integrated by specific Artificial Intelligence (AI) models for fast and accurate automatic anomaly detection. The mDT is the digital copy of the examined device in this case, it can be seen as an emulator that receives real-time data from the physical device. Therefore, it replicates all its behavior but with super-user permissions. 
In the case of an incident in industries or critical infrastructures based on mobile devices, it is fundamental to create the DT systems before the device is started to be used. On the other hand, in case of private investigations where the owner is willing to collaborate, a single live mDT can be created for live examination. 
In particular, the methodology shown in Figure~\ref{fig:methodology_detail}, follows the three main stages of a security threat: \emph{(i)} \textit{before}, when the environment is clean and no anomalies happened; \emph{(ii)} \textit{during}, while the threat is running to monitor the system and collect information; and \emph{(iii)} \textit{after}, to reconstruct the threat pipeline. This continuous monitoring is fundamental to acquire evidence of each phase, at any time and event, allowing a complete overview of the attack. Moreover, as the mDT is based on emulators that typically run on Virtual Machines (VM), it is easy to restore its saved state and recover data from a specific moment or event. 
An AI-supported mDT can be assigned to each phase, communicating via TCP/IP, to comply with the bidirectionality introduced by Grieves \etal \cite{Grieves16_DT}. In the case of a cyber attack, this proposed system can then \emph{(i)} \textit{prevent} future incidents if based on known patterns and signatures; \emph{(ii)} \textit{detect} attacks in real-time and collect forensics information; and \emph{(iii)} \textit{reconstruct} the incident and understand whether it depends on the SwA behavior or user input. In the case of a private investigation, stages for \textit{live} and \textit{post-mortem} analysis can be simplified to acquire and analyze data from the mDT as if it were the real device. We better detail this scenario in the Android case study~\ref{sec:chat}.

The \textbf{main contribution} of this paper is thus the proposal of a new mobile live intelligent forensics methodology to acquire data that would typically require super-user access. By having a remote backup copy live synchronized (mDT), data can be accessed on real time with user or super-user privileges. 
At the same time the methodology allows to monitor the device to prevent the failure of a mobile device, for example, in critical infrastructures. In this way, an anomaly is prevented. In case of a technical failure, the mDT can help with incident reconstruction. Another significant contribution is applying different technologies in the mobile forensics field to overcome the problem of super-user permission requests. The mDT allows the analysis of a digital copy of the real device with the exact same content as the real physical device (usually not analyzed if there are no super-user privileges) without compromising the real device state. It is essential not to give the super-user permissions to the real device because it can be dangerous for the infrastructure, exposing the device to more security and privacy threats, as a malware or attacker immediately has super privileges to make their malicious actions. 
Moreover, this paper shows that the DF acquisition of an mDT has the same content as the real device, allowing admin data access without affecting the real device.

The remainder of this paper is structured as follows. Section~\ref{sec:sota} gives some preliminaries about mobile forensics acquisition and the super-user data. Section~\ref{sec:motivation} highlights the need to study mobile forensics acquisition. Section~\ref{sec:MoLIFE} describes the MoLIFE methodology, and Section~\ref{sec:tech} introduces all the technologies that could be integrated into the proposed MoLIFE. Section~\ref{sec:MoLIFEscenario} first describes some use case scenarios and then applies the MoLIFE to Android devices, the most used mobile device worldwide. In particular, Section~\ref{sec:chat} shows a case study on chat acquisition from mobile devices. 
The validity of the DF acquisition in Android mDT is discussed in Section~\ref{sec:discussion}. The paper ends with the highlights on the open challenges and limitations driven by some current limitations in Section~\ref{sec:challenges}, whereas Section~\ref{sec:conclusion} outlines the conclusions and future work.

\section{Background and Related Works}
\label{sec:sota}
The NIST standard defines four main stages to be followed during a general DF investigation, not only concerning mobile: \emph{(i)} \textit{collection} to seize every device to be analyzed (\ie everything with a memory, including both RAM or disk) without damaging and invalidating the evidence (\ie do not physically break it) and taking pictures to prove it; \emph{(ii)} \textit{examination} to \textit{acquire} data from memory according to its type (\ie differences between desktop and mobile, but also between OS types and different releases/versions) without damaging it and its data; \emph{(iii)} \textit{analysis} to objectively interpret the extracted data and find the important information; and \emph{(iv)} \textit{reporting} to write and deposit the report for the legal prosecution. The evidence must support every hypothesis to reconstruct what happened. This is sometimes difficult, as the evidence belongs to a single instant instead of the whole incident time, as highlighted by Pierre Margot in ``\textit{Traceology, the bedrock of forensic science and its associated semantics}" \cite{PierreMargot}. Extracted data is prone to uncertainty and needs interpretation and correlation \cite{Hargreaves_DFRWS24}.

All commercial tools for DF on mobile devices, such as \texttt{Magnet AXIOM}\footnote{\url{https://www.magnetforensics.com/}}, \texttt{Cellebrite Inseyets}\footnote{\url{https://cellebrite.com/en/cellebrite-inseyets/}} or \texttt{Oxygen}\footnote{\url{https://www.oxygenforensics.com/en/}}, integrate the acquisition, analysis, and reporting phases. Moreover, these tools can acquire the data even if the device is not rooted (details will be described later in this paragraph). Otherwise, open-source and free tools can be used, such as \texttt{adb},  \texttt{LiMe}, or \texttt{Fridump}. The choice of tools also depends on the needed acquisition: \textit{logical} using OS APIs to copy user data; \textit{filesystem} acquiring the entire file system to retrieve deleted data and metadata; \textit{physical} bit-to-bit copy of the full flash memory; \textit{volatile} acquiring the RAM in terms of the full memory or only the allocated space of a target process. Only filesystem, physical, and volatile acquisitions typically require elevated privileges. Logical ones can often work without acquiring only the non-rooted partitions, such as the sdcard, user documents, and pictures. 

First of all, \textit{rooting} or \textit{jailbreaking} is a procedure used in a digital device to enable persistent or temporary super-user (\ie administrative) privileges. By default, for security and data protection reasons, the systems do not grant administrative privileges to all users. In this way, data integrity is preserved, and if not granted by default, it can sometimes be difficult to access specific reserved information. Also, from a malware protection perspective, it is essential to deny the privilege elevation (\ie privilege escalation) by default. Despite this, different privilege-escalation techniques have been developed for the different systems, and sometimes super-user privileges are essentially required for complete monitoring, checking, and management (\ie in Android to monitor the content of the RAM or read specific directories). In iOS, the procedure is called \textit{jailbreaking} and relies on exploiting Apple's closed system and signature verification. In Android, it is referred to as \textit{rooting} and modifies system partitions or boot configurations. Still, it depends on the OS version, Google patches, and the device vendor's model, and often it also involves exploiting vulnerabilities. From a cybersecurity perspective, such exploits can be dubious, mainly if they are found on the open web, and because the super-user exposes the device and its data to additional risks (\eg data theft). However, from the DF perspective, because the device has been altered to acquire super-user privileges, its admissibility in court is unacceptable due to the violation of the evidence preservation principle. Even if a traditional user does not have the smartphone synchronized with the mDT, we demonstrate that with specific techniques detailed in Section~\ref{sec:mobileliveforen:scenario}, their smartphone can be acquired and analyzed with super-user privileges without compromising the real device. 
This is not applied everywhere, as some legislations allow root acquisition with a carefully justified documentation, for example with video and pictures, and typically the presence of all defense and prosecution analysts. In fact, some US and Australian courts have allowed, under authorization, the use of specific intrusive procedures to extract data when reasonable and necessary in the case resolution. According to the experience of different first responders and DF analysts, this is an important problem throughout Europe. As the rooting procedure does not preserve the integrity of the device~\cite{Almehmad19_ICCAIS_rooting}, when the device is not rooted, analysts can extract only accessible  user data.

Commercial forensic suites such as {\tt Magnet Axiom}, {\tt Cellebrite Inseyets}, and {\tt Oxygen} obtain super-user privileges through a layered, model-specific approach that favors low-impact channels first and escalates only as necessary. First, they attempt to leverage existing trusted connections (\eg previously authorized {\tt ADB} or {\tt iTunes} pairing records and standard encrypted backups) and vendor diagnostic or maintenance interfaces. When these are present, they can extract large amounts of data without changing the device state. If those paths are unavailable, they may deploy short-lived, vetted agents or temporarily boot trusted code into RAM (\ie a live recovery) to mount and image partitions without flashing the device. When software interfaces are insufficient, the tools will fall back on proprietary, model- and patch-level privilege-escalation techniques (\ie private exploit chains or chipset/download-mode procedures) that vendors keep confidential. As a final resort, they support hardware methods (JTAG, ISP, chip-off) to read flash memory directly. These products, therefore, combine legitimate APIs, diagnostic protocols, runtime helpers, low-level chipset access, and, where legally and technically warranted, exploits for privilege escalation. Importantly, reputable vendors log every action, validate and hash acquired images, and surface indicators of device state changes (\eg tamper or warranty flags) because many privileged methods can alter timestamps, trip vendor tamper counters, or trigger factory wipes on bootloader unlock. The practical consequence for examiners is that the exact extraction path used depends on the device model, OS build, and patch level, and differs between Android and iOS due to platform-specific protections, such as Secure Enclave/keychain and Activation Lock, on Apple devices. Hence, any elevated access must be supported by explicit legal authority, exhaustive documentation of tools/versions/commands, and hashes. Additionally, where possible, independent validation or peer review is conducted to ensure the evidentiary integrity and defensibility of the resulting images and findings.

\begin{table*}[pos=t]
\centering
\footnotesize
\begin{tabularx}{\textwidth}{@{} l Y Y p{0.26\textwidth} @{}}
\toprule
\textbf{Method} & \textbf{Description} & \textbf{Key limitation} & \textbf{Example} \\
\midrule
Logical / backups
& Extract user data via trusted APIs or pairings
& Misses protected/system data and volatile memory; encrypted backups need credentials
& ADB, iTunes/iCloud; Feng~\cite{Feng18_DI}, Geus~\cite{geus2024systematic}, LiFE~\cite{life2017ios} \\

Agent / instrumentation
& Inject a short-lived helper into an SwA/process
& SwA-scoped; requires repackaging; limited coverage
& Frida/Fridump; Bellizzi~\cite{Bellizzi_21arxiv,Bellizzi_SIP21} \\

Temporary live-boot
& Boot a RAM-based recovery/ramdisk to mount and image partitions
& Requires unlocked bootloader or exploit; may trigger wipes
& Magnet, Cellebrite, Oxygen; Kumar~\cite{kumar2025comparative}, Ocen~\cite{ocen2024multiprocess} \\

Hardware extraction
& Read flash via JTAG/ISP/chip-off to bypass OS protections
& Expensive and potentially destructive; at-rest encryption can block recovery
& JTAG/ISP; Thing~\cite{thing2010live}, Ocen~\cite{ocen2024multiprocess} \\

\bottomrule
\end{tabularx}
\caption{Compact summary of non-persistent bypass methods, key limitations and representative examples}
\label{tab:sota}
\end{table*}

Recent academic research has increasingly explored data extraction techniques that do not require rooting or jailbreaking, mainly through logical, backup-based, and agent-assisted approaches. Because of the security protection of iOS devices, they are less analysed in the academic literature, but this does not mean they are more secure or no analysis can be made. Feng \etal \cite{Feng18_DI} proposed a logical acquisition method for unrooted Android devices. It uses system APIs to export user and SwA data without escalating privileges. Unfortunately, this method cannot access all other system or SwA directories that are protected from the super-user access or recover the deleted files. Similarly, Cuomo \etal \cite{Cuomo22_Sensors} study the importance of repeatability and evidence minimal alteration carried out by logical acquisition. While it does not affect evidence integrity, it cannot acquire volatile data, such as caches and logs. Geus \etal \cite{geus2024systematic} evaluated the forensic data acquisition using the smartphone local backup. This study reveals that while native backup techniques (\eg Android ADB backup, iTunes backup) offer a non-invasive option, their captured data is incomplete, databases are inconsistent, and the OS-version dependencies prevent reproducibility. To address encrypted devices, Fukami \etal \cite{Fukami21_FSIDI} introduced a conceptual model for forensic data extraction. The encryption boundaries can be bypassed by combining standard logical methods with vulnerability exploitation. This approach requires OS-specific vulnerabilities, hence it is not general and scalable. Kumar \etal \cite{kumar2025comparative} and Lwin \etal \cite{Lwin20_ICCA} analyze commercial forensic tools (\ie Cellebrite, Magnet AXIOM, Belkasoft, Oxygen) on unrooted Android devices. Tool-based logical acquisitions are valuable for retrieving contacts, messages, media, and specific SwA data. In contrast, they are ineffective when accessing protected or deleted content, especially before Android $11$. LiFE \cite{life2017ios} introduced an open-source logical backup examination tool for iOS, leveraging iTunes backups for artifact analysis. Many third-party SwA containers and encrypted keychain entries are not retrieved because data examination is based on what is found in the Apple backup. Most modern non-root acquisitions rely on system APIs, backups, or MDM/agent-based methods that minimize device alteration but trade completeness for safety \cite{aguirregomezcorta2025android, ocen2024multiprocess}. Bays and Karabiyik \cite{bays2019forensic}  demonstrated that SwA-specific artifacts (\eg location-sharing) can often be retrieved through logical or backup acquisition. Nonetheless, deep system data are inaccessible. Earlier works, such as Vrizlynn L.L.  Thing \etal \cite{thing2010live}, acquire RAM without rooting. Still, such techniques are obsolete due to modern OS security hardening (SELinux, verified boot). Bellizzi \etal \cite {Bellizzi_21arxiv, Bellizzi_SIP21, Bellizzi_Access22, Bellizzi_JCP23_vedrando} conducted different studies on how to acquire Android RAM without rooting. However, this methodology is strictly dependent on reverse engineering of the SwA (\ie not scalable in large-scale analysis and not generalizable because of different SwA behavior). Moreover, because it relies on Frida Gadget, the acquisition is limited to specific portions of the RAM space allocated to the target application, rather than the entire RAM. 

Collectively, these studies show that while non-root and non-jailbreak acquisition methods preserve forensic integrity, they are inherently limited in scope (\ie unable to recover deleted data, encrypted containers, or privileged system logs). Therefore, they must be complemented by privileged, agent-based, or exploit-assisted approaches when complete solutions are required.

\section{Motivation on Mobile Forensics Acquisition}
\label{sec:motivation}
Mobile forensics increasingly struggles with modern security protections on smartphones (\eg verified boot, strong encryption, hardware-backed key storage). Rooting or jailbreaking can grant full access, but these procedures are device-dependent, unreliable, and often destructive: unlocking or flashing custom firmware may wipe data or modify system state, making the evidence inadmissible. Conversely, safer approaches that avoid privilege escalation (logical, backup-based, or agent-assisted acquisitions) provide only partial visibility and exclude protected directories, encrypted containers, deleted content, and volatile memory. Commercial forensic suites can sometimes extract data from non-rooted devices using proprietary exploits or temporary agents, but these techniques operate as black boxes, offering limited transparency and poor explainability, an issue that weakens trust and defensibility in court. Overall, the field faces a persistent trade-off: full access risks damaging evidence, while safe methods lack completeness. Platform fragmentation and rapid security patching further reduce reproducibility and consistency across devices.

To overcome this, we target the core challenge of mobile acquisition: obtaining privileged-level data without altering the original device. The proposed solution, a mobile Digital Twin (mDT), creates a synchronized replica of the smartphone where elevated acquisitions and dynamic analysis can be performed. The mDT mirrors the device’s state, allowing access to filesystem, volatile memory, and kernel artifacts as if the device were rooted, while the physical phone remains untouched. Integrated into the MoLIFE methodology, the mDT enables continuous monitoring, timeline reconstruction, and forensically defensible analysis in authorized investigations, ensuring completeness without compromising evidence integrity.

\section{Proposed Methodology - MoLIFE}
\label{sec:MoLIFE}
This section presents the detailed methodology to be applied in industries and critical infrastructures to monitor continuously from day zero to the incident, a relevant mobile device. Such MoLIFE infrastructure monitors the \emph{(i)} \textit{forensics prevention} stage (cf. Section~\ref{sec:methods:preventiveforen}) to avoid the installation of specific SwA possibly associated with known threats and attacks; \emph{(ii)} \textit{live forensics} (cf. Section~\ref{sec:methods:liveforen}) to detect at real-time the attack and collect forensics data; and \emph{(iii)} \textit{post-mortem forensics} (cf. Section~\ref{sec:methods:postmortemforen}) to reconstruct the steps that lead to the incident starting from the forensics data acquired in the previous stages. The full procedure is described in the Diagram Flow depicted in Figure~\ref{fig:flowdiagram}. MoLIFE follows the main pillars of DF designed in \cite{Ayers14_NIST}, and each stage will respect them as shown in Table 2.

\subsection{MoLIFE Preamble}
\label{sec:methods:preamble}
Each stage of MoLIFE is managed by a mDT, an emulated and virtualized mobile device with super-user privileges, interacting with the real device, having the same hardware characteristics and software logics (except a few cases as highlighted in Section~\ref{sec:mobileliveforen:androdt}, but leading to the same working mechanism and final results). The mDT, as a traditional DT, receives data from the real device in real-time (complete synchronization), processes them, and sends feedback to the real mobile device (bidirectionality principle). The communication is possible thanks to WiFi debugging, with the condition of having the mDTs and the real mobile device in the same network, for example, using a Virtual Private Network (VPN). Data elaboration on the mDT can be integrated with AI algorithms to speed up the decisions, making them more accurate, described in the following.

\subsubsection{\textbf{Digital Twin}}
\label{sec:dt}
Getting inspired by the traditional DT (\ie  a synchronized virtual replica of a physical system that processes data in real time, enabling monitoring, simulation, and analysis, often supported by AI, widely used in industry, critical infrastructures, IoT, and IIoT\cite{grieves_digital_2016, boschert_digital_2016,liu_review_2021,Fuller20_Access,maddikunta_industry_2022}), the concept of mDT (\ie a DT of a mobile device) is introduced. The mDT can be applied in mobile DF investigations in \emph{(i)} \textit{acquiring data} from the infrastructure by solving the super-user problem related to data loss (\ie elevating the privileges before the synchronization); \emph{(ii)} \textit{preventing and detecting} real-time incidents by using highly computational costing algorithms; and \emph{(iii)} \textit{granting a backup and a collection} of detailed and specific data for each time in case of a future forensics analysis to understand the incident source. The mDT, being a synchronized replica of the real device, not only contains personal data but may also embed intellectual property, such as proprietary configurations, business applications, or sensitive corporate assets. 
Nowadays, there is no DT for mobile devices, but some emulators have been released (\eg Android Studio, Genymotion, AWS, XCode, etc.), allowing to run tests on a virtualized mobile device without the need of a physical mobile device. To turn these emulators into pure mDTs, data must be synchronized between the emulator and the real device through bidirectional communication. They should be in the same state simultaneously (or with a delay of very few seconds due to the latency), as in the IIoT. In the case of mobile devices, the synchronization could be done in two main ways: \emph{(i)} using the WiFi debugging on the emulator and the real device, allowing to send and receive data over TCP/IP; and \emph{(ii)} accessing to the same target SwA in the emulator and the real device with the same credentials and having data automatically synchronized via an external server managing the SwA (\eg messaging apps, social networks and all the SwA with a cloud behind). The first presented solution is not that easy to implement. First of all, the input from the target SwA in the real device must be intercepted (\ie needing an interceptor, such as {\tt Frida}\footnote{\url{https://frida.re/}}, to hook into a specific function of the target SwA and capture the input). 
Secondly, the input must be sent to the mDT via TCP/IP. {\tt Frida} is a dynamic instrumentation toolkit for injecting scripts into running processes, enabling real-time analysis and manipulation across platforms. Additionally, it is necessary to modify the execution of the target SwA in the mDT to receive and process the captured input. At the same time, the execution of the target SwA must be stopped both in the real device and in the mDT and restored once the input is processed. The procedure is challenging because it causes a loss of time, meaning a desynchronization. Moreover, this technique needs a repacking on the target SwA, creating two versions from the original: \emph{(i)} one in the real device for the non-rooted interception mechanism with \textit{frida-gadget}, and \emph{(ii)} the second for the mDT to receive the input properly. Repacking a SwA is sometimes impossible due to anti-tampering techniques to prevent reverse-engineering and software cloning \cite{Guo20_LNCS}. In some cases, such as for memory forensics, data cannot be injected in the RAM. The same holds for permanent storage, which can be simply copied and pasted. For that reason, it should be better to have the same access of the target SwA in the mDT and the real device, allowing automatic data synchronization, only focusing on behavior interception on the mDT and sending a stop command with the WiFi debugging to the real device in case of a failure. 
The use of an mDT is more helpful than the analysis of the mobile data stored in the cloud because it allows to \emph{(i)} \textit{analyze the full SwA} behavior and not only dump its content; \emph{(ii)} \textit{analyze the effects} of the target SwA in the system; \emph{(iii)} \textit{monitor the target SwA}, above all even without root access; and \emph{(iv)} \textit{access its data immediately } without waiting another jurisdiction to answer and granting the analysis of the data stored in the cloud of another country with different legislation and data management. Additionally, the mDT helps adding computationally expensive tasks that the real mobile device could not support. An example is the case with AI detection models or other identification  that in a real device could not be supported because of resource draining (\eg battery, CPU, storage, RAM, temperature). 

In summary, \textit{what is the main difference between an mDT and an emulator?} An mDT is a real-time virtual copy of a physical device that stays continuously synchronized with it. Unlike a typical emulator, which only simulates a generic environment, the mDT mirrors the actual device’s current state and behavior, enabling forensic analysis, monitoring, and incident reconstruction without modifying the physical evidence.

\subsubsection{\textbf{Artificial Intelligence}}
\label{sec:ai}
The mDT is designed to support analysts in detecting anomalous behaviors and potential threats that may indicate security incidents or malicious activities. Such detection requires identifying deviations from normal usage patterns, which is inherently complex given the diversity of user behaviors and the volume of data produced by mobile devices. To address this challenge, AI algorithms can be easily integrated, as they allow the recognition of subtle anomalies, accurate classification of suspicious events, and prioritization of alerts. In particular, semi-supervised Deep Learning (DL) models are used to overcome the lack of labeled training data. However, the adoption of AI-based approaches introduces the well-known challenge of interpretability: complex models such as DL often act as “black boxes,” making it difficult for human analysts to justify or validate their outcomes. In forensic investigations, this limitation is critical, as the reliability and admissibility of evidence depend on the analyst’s ability to explain how a decision was reached. For this reason, eXplainable AI (xAI) techniques are integrated into the mDT, making the decisions of AI models interpretable by highlighting which features (e.g., specific API calls, anomalous network traffic, or user behavior patterns) influenced the classification. This ensures that analysts can understand and defend the results, strengthening their trustworthiness in judicial or corporate contexts. Additionally, the system can integrate Cyber Threat Intelligence (CTI) feeds, which provide external contextual knowledge (\eg known indicators of compromise, emerging attack patterns). These CTI sources are processed using Large Language Models (LLM) and Natural Language Processing (NLP) algorithms, which enable automatic extraction of key information, summarization of unstructured threat reports, and identification of relevant patterns at scale. Beyond detection, the same algorithms support analysts in generating clear, human-readable reports for both preventive and post-incident analysis. Since these models require significant computational power and energy, they must be executed on the mDT hosted on a server with higher resources but synchronized with the mobile OS, ensuring accuracy and real-time data without draining the device’s resources.

\begin{figure}[pos=t]
    \centering
    \includegraphics[width=\linewidth]{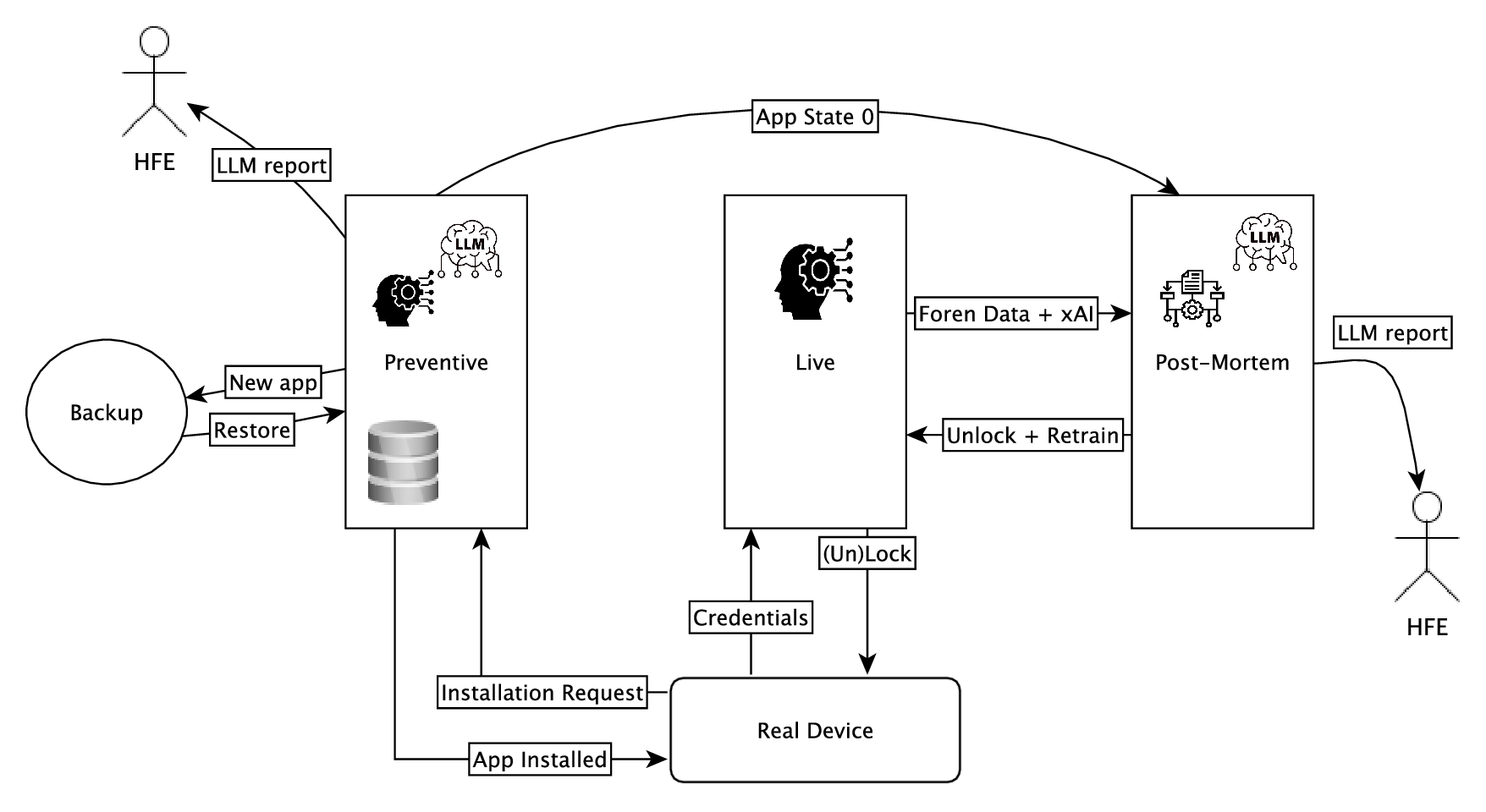}
    \caption{The methodology operates in three coordinated stages: Preventive, Live, and Post-Mortem. The Preventive stage analyzes installation requests using a clean backup; if safe, it installs the app. If an anomaly is detected, control moves to the Live stage, which locks the real device and collects forensic and xAI data. The Post-Mortem stage analyzes this data, reconstructs events, and, if no threat is confirmed, instructs the Live stage to unlock the device. This process generates a complete timeline of the incident}
    \label{fig:methodology_detail}
\end{figure}

\subsection{Stage 1: Preventive Forensics}
\label{sec:methods:preventiveforen}
This stage is necessary to avoid installing a specific SwA that exhibits characteristics or behaviors commonly associated with known malware or potential threats. As the name suggests, it is designed to prevent a cyber incident. When the user wants to install a new SwA in the real device, it sends a request to the mDT associated with this stage. Thus, the mDT analyzes the SwA accordingly and, if no anomaly is detected, the SwA can be installed on the device. 

The analysis is mainly made up of two stages. A first one is made on the \textit{\textbf{hash matching}} to prevent the installation of an unknown and unofficial SwA. A SwA could be repacked to add exploitation or malware behavior, or retrieve important owner data and change the apparently official SwA \cite{Soi25_itasec}. To ensure the integrity of the SwA, the first step is to verify its hash against possible repacking or tampering~\cite{Chen20_TIFS, repack_ios2}, where a legitimate application may have been modified to include hidden malicious code~\cite{Soi25_itasec}. 
The modified SwA has a different hash than the original one but no match with ssdeep, feature hash, permhash. The hash detector verifies whether a SwA is unofficial or repacked by comparing it against a database of known legitimate and previous app versions, while code similarity techniques can further aid in detecting potential malicious modifications \cite{Ahammed24_GSC}.  
Different CTI online systems can be queried on the computed SwA's hash to improve security. These systems (\eg Virus Total, MITRE CVE, NIST NVD),
have a report and information on many SwA stored by hash. Besides hash-based verification of SwAs, the database can be enriched with additional data from CTI sources, such as malicious URLs, domains, or phone numbers used in phishing. In this context, LLM and NLP algorithms are useful to automatically analyze unstructured CTI reports or online data, extract relevant indicators of compromise, and update the database with actionable entries for MoLIFE. Additionally, LLM and NLP algorithms can automatically give a comprehensive report to the analyst. 

If the check of the hash value is passed, the SwA is analyzed by a pre-trained AI model to detect specific static and dynamic features related to vulnerabilities and malicious behaviors \cite{Kambar22_CCWC}. Static analysis inspects the code without execution, highlighting suspicious elements such as API calls, cleartext strings, or IP addresses, while dynamic analysis observes the execution of the SwA in a sandbox, monitoring function calls, memory usage, and network traffic in a safe and controlled environment. AI enhances these traditional approaches by learning complex patterns from both static and dynamic data, improving the recognition of known malware families and their evolving variants. However, the use of AI in this context must also consider the risk of adversarial attacks, where small crafted modifications are introduced to mislead the detection model and induce the misclassification of a malicious sample as benign; this highlights the importance of robust feature selection and resilient training strategies. When no threat is detected, the SwA is first installed in the mDT for live forensics analysis and subsequently on the real device; if instead the algorithm raises an alarm, installation is blocked, a report is automatically generated for the human operator, and the environment is restored to a clean state to remove any persistence mechanisms. Finally, to make the AI decision interpretable and to justify why the SwA has been classified as “malicious” or “benign,” xAI techniques (\eg feature attribution methods such as SHAP or LIME) are applied, allowing the analyst to understand which features influenced the classification and to validate the reliability of the decision \cite{Soi24_JISA}.

According to NIST SP800-101, this stage follows all the phases: the \textit{\emph{(i)}} \textit{collection} phase to identify data from the running application; the \emph{(ii)} \textit{examination} because it extracts potentially risky features from the application; regarding the \emph{(iii)} \textit{analysis} phase, the system automatically interprets and takes a decision on the extracted data with a conjunction analysis from external CTI tools and in case of a detected suspicious behaviors, \emph{(iv)} \textit{report} it to the human analyst.

\begin{algorithm}[t]
\caption{MoLIFE Prevention Algorithm}
\label{alg:MoLIFE-prev}
\begin{algorithmic}[1]
\Require SwA, hash database, CTI reports
\Ensure Pre-trained models on static and dynamic features
\State Hash Match
\State CTI-driven Analysis
\State Extract static and dynamic features
\State \textbf{Return}: Classification
\State \textbf{Output}: Report with LLM
\end{algorithmic}
\end{algorithm}

\subsection{Stage 2: Live Forensics}
\label{sec:methods:liveforen}
Malware sometimes reveals its malicious behavior after an elapsed time or event, or when a vulnerability is exploited or by specific user input \cite{Fratantonio16_SSP, Samhi22_TDSC}. Moreover, the analysis must be fast, since a SwA can only remain in “quarantine” for a short time before the user needs to use it, especially in critical scenarios. 
For this reason, the preventive forensics stage is not sufficient in itself. 
Accordingly, MoLIFE needs live forensics analysis to continuously monitor the SwA and its effects on the device at runtime and during standard execution. 
The live forensics stage supervises and collects forensics data in case of future possible DF analysis with a more forensically oriented algorithm. For example, the full RAM content can be analyzed to check the effects of a specific target SwA in the whole system at runtime. In addition, the target SwA can be inspected with instrumentation tools (\eg Frida) to check specific behaviors at run-time. The network connections and communications can be inspected to check which IPs and domains the SwA contacts and retrieve the communication payload. As the mDT has super-user privileges, specific directories can be monitored to check if uncommon and potentially dangerous files are downloaded, created, or accessed by the SwA at runtime and stored in the device. It is worth pointing out that this simulation technology is potentially dangerous, as the mDT includes intellectual property and personal data, whose disclosure could cause significant consequences and irreparable damages \cite{Kuštelega24_JUCS, Sun15_ACM}. 
In this scenario, the mDT can check if specific, unauthorized, illegitimate, and uncommon data is read and exploited by the SwA outside of its classic, standard, and declared behavior obtained from the previous preventive forensics stage investigation.
The \textit{\textbf{forensics-based AI model}} used must be robust to adversarial attacks to avoid an attacker using anti-forensics techniques \cite{Mohammad24_ICSPIS, Kessler2007_ADFC} and steganography (\ie hiding information in other sources such as images, audio, video, text, to not be easily detected) techniques \cite{Soi25_itasec, badhani2018evading} to hide behaviors and bypass detection. Even at this stage, the AI algorithm must include xAI-based techniques to find the features that have the highest influence in classifying the behavior as a security threat, and the related reason. This is important for the next post-mortem analysis stage to reconstruct what happened. In case of a cyber attack, the SwA is blocked first on the real device to prevent more damage, avoiding desynchronization (\ie the mDT must have all data of the real device), and, lately, on the mDT. Subsequently, the collected data are sent to the post-mortem forensics stage to reconstruct what happened. 

This stage also considers the NIST SP800-101 in the \emph{(i)} \textit{collection} phase when it identifies the forensics data to which to pay attention during the SwA execution; the \emph{(ii)} \textit{examination} phase by extracting information and features from the identified forensics data; the \emph{(iii)} \textit{analysis} to interpret the forensics data under attention. The \textit{Live Forensics} stage does not have a traditional \emph{(iv)} \textit{reporting} phase because no final report is generated for the human analyst, but in some sense, sending the features that determined the threat classification and the collected data to the next stage could be considered a sort of report as the information is described.

\begin{algorithm}[t]
\caption{MoLIFE Live Algorithm}
\label{alg:MoLIFE-live}
\begin{algorithmic}[1]
\Require SwA
\Ensure Pre-trained models, live forensics dynamic features
\State Network Traffic Analysis
\State RAM Monitoring
\State Disk File Analysis (content and creation)
\State Explain the classification (xAI)
\State \textbf{Return}: Live Detection Classification
\State \textbf{Output}: xAI, collected forensics data
\end{algorithmic}
\end{algorithm}

\subsection{Stage 3: Post-Mortem Forensics}
\label{sec:methods:postmortemforen}
To understand whether the threat detected has been raised by user input or hidden behavior of the SwA, a third stage is required to investigate what happened, \ie post-mortem analysis. To start the analysis, the post-mortem forensics stage needs the original data at the time of installation from the preventive stage (\ie SwA in the \textit{state$_{0}$} after the analysis and before installation). Additionally, it receives from the live stage its conducted analysis: \ie \emph{(i)} the incident information with the actual state of the SwA, \emph{(ii)} the forensics data and \emph{(iii)} the xAI features that lead to the alarm of a suspicious behavior. At this stage, the mDT continuously sends specific inputs to the original SwA and stops when it reaches the state of the SwA at the moment of the attack (\ie the state of the SwA recovered from the live forensics stage).  
These capabilities are implemented using \textbf{fuzzing} techniques to trigger and detect vulnerabilities, bugs, or anomalous behavior \cite{wong_intellidroid_2016, kroll_aristoteles_2021}.
Random input is generated from a starting set, usually defined by the user according to the specific SwA. Recent research has begun to explore the integration of AI into fuzzing, with LLM showing particular promise \cite{xia_fuzz4all_2024}. Traditional fuzzing approaches typically rely on random or heuristically generated inputs, which may fail to capture the structure or semantics of the target program. In contrast, LLM can synthesize syntactically valid and semantically rich test cases that are tailored to the program’s logic, increasing the likelihood of exposing subtle bugs, vulnerabilities, or anomalous behaviors. Although still an emerging direction rather than a mature technology, this line of work demonstrates how generative models can extend classical fuzzing by producing more realistic and context-aware input variations, thereby improving both coverage and effectiveness. The program is executed each time with a single input, and its behavior and output are tracked. No vulnerability or security issue is reported if the output is the same as the expected one. In contrast, if some crash, error, or anomaly is detected, the fuzzing program reports it to the human analyst, logging the input that produced it. 
In the literature, there are three different types of fuzzing: \emph{(i)} \textit{black-box}, when the tests are made on external input and the fuzzer does not know the program's structure under test, and general analysis is made; \emph{(ii)} \textit{white-box}, fuzzing when the program source code and structure are available, \ie specific test cases are generated, and the analysis is specific; and \emph{(iii)} \textit{gray-box}, fuzzing to use a restricted set of internal program's knowledge for the tests. 

Fuzzing is used in MoLIFE to determine what led to the threat and reconstruct how it happened. In particular, if the threat has been raised by user input (and which one), or if the real hidden behavior of the SwA emerged after a specific elapsed time or event.  
As the structure of the target SwA is known from the analysis conducted in the preventive stage, white-box fuzzing is preferred with inputs generated by specific LLM for wider coverage. Moreover, the fuzzing procedure can be led by an AI model that previously explored the structure of the target SwA from the information collected in the preventive forensics stage and its prior knowledge. At this stage, as for preventive forensics, the data collected for the analysis are also analyzed through online and reliable CTI tools. Their output is processed via LLM, not influencing the algorithm's decision but just giving more details and data on which to base the output; they are only additional information to enrich the decision and improve the final report. In the case the fuzzing algorithm does not detect the threat, this means that the AI model in the live forensics stage made some error (and this can happen because of insufficient training data or for anti-forensics, steganography, anti-analysis techniques used by the malware \cite{Alrammal22_JIT}). Hence, the mDT will send a notification to continue the execution first on the live forensics mDT and then on the real device (because of the synchronization paradigm) after the AI model of the live stage is retrained. Otherwise, the same incident is reported. On the other hand, if a cyber-attack has been detected, the execution remains blocked. Subsequently, a report with LLM is generated for the external forensics human analyst to know \emph{(i)} what happened, \emph{(ii)} make a manual analysis, \emph{(iii)} and make a correct decision. 

\begin{algorithm}[!t]
\caption{MoLIFE Post-Mortem Algorithm}
\label{alg:MoLIFE-post}
\begin{algorithmic}[1]

\Require SwA in $state_0$, Live Features, xAI Live Analysis, CTI reports
\Ensure Pre-trained AI-based fuzzing
\State Fuzz the SwA in $state_0$
\State Stop when reaching the same forensics data as the Live Analysis
\State \textbf{Return}: Decision on retraining previous algorithms based on new features and findings
\State \textbf{Output}: Report the incident's causes

\end{algorithmic}
\end{algorithm}

This stage also contemplates the NIST SP800-101 in the \emph{(i)} \textit{collection} because it receives the data from the preventive forensics stage (\ie the original SwA and its analysis), from the live forensics system (\ie the state of the SwA at the moment of the incident, the xAI features) and from external CTI tools. The \emph{(ii)} \textit{examination} phase is used to interpret all the received data, the original SwA, and the incident state. The \emph{(iii)} \textit{analysis} phase corresponds to the check at each step of what happened and how the incident occurred by using the fuzzing techniques. When the incident state is reached, or after many analyses, no threat is detected, the system gives in output a \emph{(iv)} \textit{report} of the findings to the human forensics analyst.

\section{Support Technologies for MoLIFE}
\label{sec:tech}
MoLIFE system, despicted in Figure~\ref{fig:methodology_detail}, may need some additional technologies to work: a communication system, AI algorithms, and a powerful device (\eg a server) to run the emulator for the mDT. For this reason, this Section highlights how the technologies can be applied to MoLIFE. The selected technologies must comply with the main cybersecurity principles. We selected the most important for MoLIFE, \ie the minimum requirements to be followed: \emph{(i)} the \textbf{CIA-triad} \ie limiting data access (confidentiality), ensures accurate
data (integrity) and makes them always accessible (availability); \emph{(ii)} \textbf{resilience} to cyber attacks, robustness, avoiding a failure and
collapse of the system; and \emph{(iii)} \textbf{non-repudiation} 
integrity and authenticity. They are represented in Table 2.

Additionally, MoLIFE needs \emph{(iv)} fast \textbf{communication} system to ensure data synchronization between the physical device and the emulated one. Cloud, Edge, Hybrid, and Quantum Computing ensure this. The first solution offers high storage and processing power but requires high energy consumption and high latency when accessing the data. For this last reason, the Cloud is inefficient in some scenarios, and hence, the Edge solution is preferred as it ensures the nodes' high mobility is closer to the end user. An intermediate solution is the use of Hybrid Computing, \ie a combination of multiple types of computing systems or architectures to optimize performance, efficiency, and flexibility. Hence, Hybrid Computing allows organizations to combine the scalability of the public Cloud with the security and control of private infrastructure, enabling flexible, efficient, and secure computing environments. Quantum Computing leverages the principles of quantum mechanics (\eg superposition and entanglement) to solve complex problems significantly faster than classical computing systems. It allows fast communication, fast data processing, and better performance of AI algorithms. Accordingly, quantum computing is a good option because when it is available, we will have the fastest executions. However, at the time of writing this article, the technology is not yet ready for practical deployment, although promising studies suggest it may become available in the future \cite{khan24NP}. For the mDT, fast and reliable communication already relies on the integration of established mobile technologies (\eg WiFi, Bluetooth, 5/6G), which are sufficient to support data synchronization. While quantum technologies may eventually find applications in domains where extremely high computational performance is required, their adoption is expected to be limited to specialized contexts and is not a prerequisite for the feasibility of the mDT proposed here. 

\begin{table*}[pos=t]
\centering
\footnotesize
\setlength{\tabcolsep}{6pt}
\renewcommand{\arraystretch}{1.15}

\resizebox{\textwidth}{!}{%
\begin{tabular}{llccc c| c ccc c c}
\toprule
& & \multicolumn{3}{c}{\textbf{Stages}} & & \multicolumn{6}{c}{\textbf{Requirements}} \\ 
\cmidrule(lr){3-5} \cmidrule(lr){7-12}
&& \textbf{Preventive} & \textbf{Live} & \textbf{Post-Mortem} & & \textbf{Communication} & \multicolumn{3}{c}{\textbf{CIA}} & \textbf{Resilience} & \textbf{Non-Repudiation} \\
\cmidrule(lr){8-10}
&&&&&& & \textbf{Confidentiality} & \textbf{Integrity} & \textbf{Availability} & & \\
\midrule

\rowcolor{gray!20}%
\multirow{8}{*}[-1em]{\begin{sideways}\textbf{Technologies}\end{sideways}}%
& mDT & $\checkmark$ & $\checkmark$ & $\checkmark$ & & $\circ$ & $\circ$ & $\circ$ & $\times$ & $\circ$ & $\times$ \\
& AI algorithms & $\checkmark$ & $\checkmark$ & $\checkmark$ & & $\circ$ & $-$ & $\circ$ & $\circ$ & $\circ$ & $-$ \\
& Mobile Communication & $\checkmark$ & $\checkmark$ & $\checkmark$ & & $\circ$ & $\circ$ & $\circ$ & $\circ$ & $\circ$ & $\checkmark$ \\
& Cloud Computing & $\checkmark$ & $-$ & $\checkmark$ & & $\times$ & $\circ$ & $\circ$ & $\times$ & $\circ$ & $\circ$ \\
& Edge Computing & $\checkmark$ & $\checkmark$ & $\checkmark$ & & $\checkmark$ & $\circ$ & $\circ$ & $\checkmark$ & $\circ$ & $\circ$ \\
& Hybrid Computing & $\checkmark$ & $\checkmark$ & $\checkmark$ & & $\checkmark$ & $\circ$ & $\circ$ & $\checkmark$ & $\circ$ & $\circ$ \\
& Blockchain & $\checkmark$ & $\checkmark$ & $\checkmark$ & & $-$ & $\checkmark$ & $\checkmark$ & $\checkmark$ & $\checkmark$ & $\checkmark$ \\
& Quantum Computing & $\checkmark$ & $\checkmark$ & $\checkmark$ & & $\checkmark$ & $\circ$ & $\circ$ & $\checkmark$ & $\circ$ & $\circ$ \\
\midrule
\multirow{4}{*}{\rule{0pt}{2em}\rotatebox[origin=c]{90}{\textbf{NIST DF}}}
& Collection & $\checkmark$ & $\checkmark$ & $\checkmark$ & & $-$ & $-$ & $-$ & $-$ & $-$ & $-$ \\
& Examination & $\checkmark$ & $\checkmark$ & $\checkmark$ & & $-$ & $-$ & $-$ & $-$ & $-$ & $-$ \\
& Analysis & $\checkmark$ & $\checkmark$ & $\checkmark$ & & $-$ & $-$ & $-$ & $-$ & $-$ & $-$ \\
& Reporting & $\checkmark$ & $-$ & $\checkmark$ & & $-$ & $-$ & $-$ & $-$ & $-$ & $-$ \\
\bottomrule
\end{tabular}

}

\vspace{0.6ex}

\begin{minipage}{0.47\textwidth}
\raggedright
\small
\textbf{TABLE 2:} MoLIFE stages, technologies, forensics. The table shows the
technologies adopted for each MoLIFE stage and their security requirements.
The last lines of the table show the 4 main principles of DF designed by NIST
and respected in the 3 MoLIFE stages
\label{tab:tech_nist}
\end{minipage}
\hfill
\begin{minipage}{0.50\textwidth}
\small
\textbf{Legend}\\[-2pt]
\begin{tabular}{@{}ll@{}}
$\checkmark$ & Satisfies by design (essential requirement) \\
$-$          & Dissatisfies by design (not needed) \\
$\times$     & Dissatisfies by design (essential requirement) \\
$\circ$      & Requires proper setup (essential requirement) \\
\end{tabular}

\vspace{2pt}
\colorbox{gray!20}{\parbox{\linewidth}{\centering \textbf{MoLIFE main essential technologies}}}
\end{minipage}

\end{table*}

As MoLIFE must comply with the CIA-triad, the use of blockchain technology can help. Blockchain technology is a decentralized distributed digital ledger that records data exchange (\ie transaction) between multiple entities to ensure security, transparency, and immutability. Generally, it is used for cryptocurrency transactions~\cite{Alcaraz2020a}, smart contracts, supply chain management, and voting systems. Blockchain technology tracks every data exchange between nodes by giving a unique hash to the nodes and transactions. Therefore, blockchain technology can track data inconsistency, maintaining the principle of data integrity in Digital Forensics.
Hence, by design, the blockchain technology proves data integrity and non-modification, a fundamental principle of MoLIFE. Other systems, such as mobile communication systems, AI algorithms, mDT, and computing systems, must be implemented correctly to ensure data integrity. For example, the communication system must not lose any transmitted data and guarantee that no one can intercept and modify it, \ie resilient to sniffing attacks. Moreover AI must guarantee data integrity with an accurate and consistent prediction, \ie samples are correctly classified and do not change with the different systems used. At the same time, data integrity is ensured with mDT and computing systems by ensuring that no one can change the stored data. For example, different solutions such as anti-viruses, access control, action logging, and firewalls can be implemented.

Due to data decentralization, the blockchain guarantees data availability every time. By design, data availability is also ensured with Edge and Hybrid Computing by combining the strengths of local processing with the scalability of cloud resources. Quantum computing does not directly guarantee data availability, but it has enormous potential to optimize, secure, and support the infrastructure that will ensure data availability in the future. Data availability in AI refers to ensuring that high-quality, relevant, and timely data is accessible whenever AI systems need it for training, inference, or continuous learning. Without strong data availability, AI systems degrade in performance, become biased, or even fail outright. Hence, AI algorithms must be implemented in this way. A good implementation of Mobile Communication systems with 5/6G, WiFi, and Bluetooth can guarantee data availability, \ie data can be transmitted, accessed, and received at any time. For example, implementing network redundancy, adaptive routing, load balancing, and Quality of Service mechanisms.

The blockchain guarantees confidentiality by default thanks to the encryption implemented in the stored chain, and only parties with the decryption keys can read the data. Due to the use of smart contracts and trusted execution environments, data is processed privately. Other systems can guarantee the confidentiality principle with a correct implementation. For example, in AI, the model must not use private data for the training; they must be anonymized, and no one can infer the training dataset in any way. This also means that the mDT and computing systems must be implemented with access control logic to ensure data confidentiality and possible modifications. 

The CIA-triad is strictly related to the resilience principle. MoLIFE and its supported technologies must be designed to be robust to cyberattacks. By design, it is ensured by the consensus mechanism implemented in the blockchain and its decentralization design (\ie without a central point of failure). The other technologies must be implemented correctly to be resilient. For example, AI algorithms must be robust to adversarial and presentation attacks so that no one can infer the model to know how it decides and modify the sample to bypass detection (\ie a malicious software not detected by the system). At the same time, the mDT and the computing systems must implement standard measures to be robust to cyberattacks, \eg robust IDS and anti-viruses. 

Non-repudiation is necessary by MoLIFE to demonstrate who accessed the forensics data, what they did, and how they modified it. This requirement is ensured by the blockchain's design, which can keep track of the transaction and data modification on the node because of the decentralized consensus, ledger immutability, and cryptography. The communication systems also ensure the principle by design because every connection is identified by specific data such as IP addresses, MAC addresses, phone numbers, and Bluetooth addresses. The AI algorithms are not designed to guarantee non-repudiation; so far, it does not depend on a good implementation. The mDT and computing systems can guarantee non-repudiation with proper access control.

\section{MoLIFE-based Scenarios}
\label{sec:MoLIFEscenario}
This section lists one of the MoLIFE possible application scenarios~\ref{sec:mobileliveforen:scenario}. 
Section~\ref{sec:mobileliveforen:androot} first gives some background on Android Forensics (the most used mobile OS worldwide) and Section~\ref{sec:mobileliveforen:androdt} presents the digital forensics acquisition from an Android mDT.

\subsection{Application Scenarios}
\label{sec:mobileliveforen:scenario}
The MoLIFE system has been designed for those specific industries classified as critical infrastructures, where it is important to protect a human operator, and specific data or areas of the industrial plant while using a smartphone device. A human worker has a specific smartphone device used for job activities and that can be used only in a particular area with a VPN. The mobile device can access specific industrial data with which the human operator works. 
Such systems are called Mobile Device Management (MDM) \cite{Timms17_CFS}. In most cases, the companies rely on third-party SwA to monitor the job-mobile devices, presuming on their implementation for security. Such SwA could contain some vulnerabilities that can be exploited if the device or the company infrastructure is not protected.  
Instead, with this system of mDTs, the protection directly depends on and relies on the company itself. In MoLIFE, the given mobile device is lightly modified to interact with the system of mDTs, \emph{(i)} \textit{prevent} the installation of specific SwA unless the preventive forensics stage sends a granting signal (\ie deny SwA installation unless authorized); \emph{(ii)} \textit{synchronize} with the mDT in the live forensics stage by sending data automatically and in live mode; and \emph{(iii)} \textit{block} a specific SwA execution and resume it when an appropriate signal is received.  
This smartphone can protect the user and the company's sensitive data (preventing industrial cyber espionage) thanks to the mDT system.

Additionally, this methodology can be adapted to other real life situations. For example, the mDT could be used to monitor in real time a victim (\eg child, woman or person abused) or target person (\ie a criminal) by having a real-time copy of the device thanks to the synchronized mDT (\ie only the live MoLIFE stage with adapted features to the examined case such as pedopornography pictures or violence words). In this way, the evidence can be acquired evidence when specific alerts are raised. For example, retrieving the content of specific chats, deleted messages, one-shot pictures as will be highlighted in Section~\ref{sec:chat}. Moreover, the mDT can be used to acquire only the current state of the device just when an accident or crime occurs without previous monitoring. This is possible only if and when the user allows for collaboration because the rooted emulator can be used as a new device where to transfer the backup of the current real device. This is the same procedure followed when changing smartphone in real life, backup all data (\ie including chats, pictures, databases etc) and transfer them to the new device. This real-time monitoring must be done according to the legislation of the Country where the victim is and a jointly jurisdictional collaboration on the Nations where they could travel. Technically, it must be supported by a continuous connectivity as the person needs to connect to the VPN where their own mDTs are stored and running on the Edge, \ie using a WiFi or $5/6$G connection. For the forensics acquisition in a violence scenario, a more specific AI system can be designed and integrated in the pipeline of this forensics system but directly on the real device used by the victim. This AI methodology will help in detecting nude pictures in the case of child abuse, child luring and child pornography or nudity for cases of revenge porn; abuse and violence messages in case of a general violence, not only for violence against women (even if it is more diffuse). A solution could be the integration of LLM and NLP on the AI detection mechanism which could work as a spam filtering, recognizing specific violence words with their synonyms or gibberish meaning in the different languages spoken worldwide. While for the media detection, the AI system will be based on object detection and image categorization. Such detection is already managed by modern DF tools such as {\tt Magnet.AI} in {\tt Magnet Axiom} which can recognize the use of violence, nudity, luring, drugs, weapons in forensically acquired data from chats and pictures, even if some studies demonstrated that they are not robust enough on presentation attacks \cite{sanna2025eusipco, Sanna24_arxiv}.

\subsection{Android Forensics}
\label{sec:mobileliveforen:androot}
Android is the most widely used mobile operating system and thus a primary target of mobile forensics. DF acquisition on Android often requires low-level operations such as unlocking the bootloader, escalating privileges, and extracting volatile and non-volatile data. The bootloader is locked by default to protect system integrity. Where permitted, it can be unlocked (e.g., via oem unlock) or bypassed using model-specific exploits, enabling the flashing of patched boot images or root managers (e.g., Magisk) to obtain superuser access. With elevated privileges, examiners can perform physical, filesystem, and volatile acquisitions: bit-for-bit imaging ({\tt dd} on block devices), full partition extraction, and RAM capture (e.g., using LiMe or fridump with analysis via Volatility/Rekall). Logical acquisitions, using OS APIs to access user-visible data, do not require root but are limited in scope. Commercial suites aim to minimize device alteration by first using trusted channels (paired backups, diagnostic interfaces) or temporary in-RAM agents, resorting to proprietary exploits or hardware methods only when necessary. All actions are logged and hashed, as privileged procedures may alter timestamps, trip tamper flags, or trigger resets. Academic work explores non-root methods (logical, backup-based, MDM/agent-assisted) that preserve integrity but cannot access deleted, encrypted, or low-level system artifacts, reinforcing the trade-off between completeness and preservation. Rooting or flashing remains intrusive and legally sensitive, requiring authorization and exhaustive documentation. To mitigate these constraints, the mDT creates a synchronized virtual replica of the device. Privileged acquisitions and dynamic analyses occur on the twin, not the physical handset, preserving evidence, maintaining a verifiable hash chain, and ensuring reproducibility. Compute-intensive AI and CTI tasks run on mDT infrastructure, avoiding disruption of the real device and concentrating sensitive processing within a controlled, auditable environment.

\subsection{DF Acquisition of Android DT}
\label{sec:mobileliveforen:androdt}

This section presents a case study on the validity of an Android forensics acquisition in an mDT. The outcome is that the forensics acquisition of the mDT produces the same result as acquiring a real rooted Android device. In detail, the section explains the different acquisition and analysis methodologies to be followed according to the data to be analyzed and the difference between the mDT and the real device. Because of the various architectures on which the real device and the mDT (some internals cannot be emulated) run, sometimes the acquisition methodology can change. As a result, the extracted content and analysis can also differ slightly. It will be detailed in Section~\ref{sec:discussion}. 
The following subsections highlight how to acquire data in the mDT, starting with the methodology used on the real device. In the following sections, we call the mDT as a mere emulator when considering only its structure and the digital copy of the real device, its emulation. Hence, it does not depend on the capability of receiving, elaborating, and sending data in real-time synchronized with the real device (the definition of the emulator being called mDT). Each acquisition is described considering both the full system acquisition or the target SwA. This second one is fundamental when dealing with privacy, preserving the user secret data not dealing with the object of the juridical process.

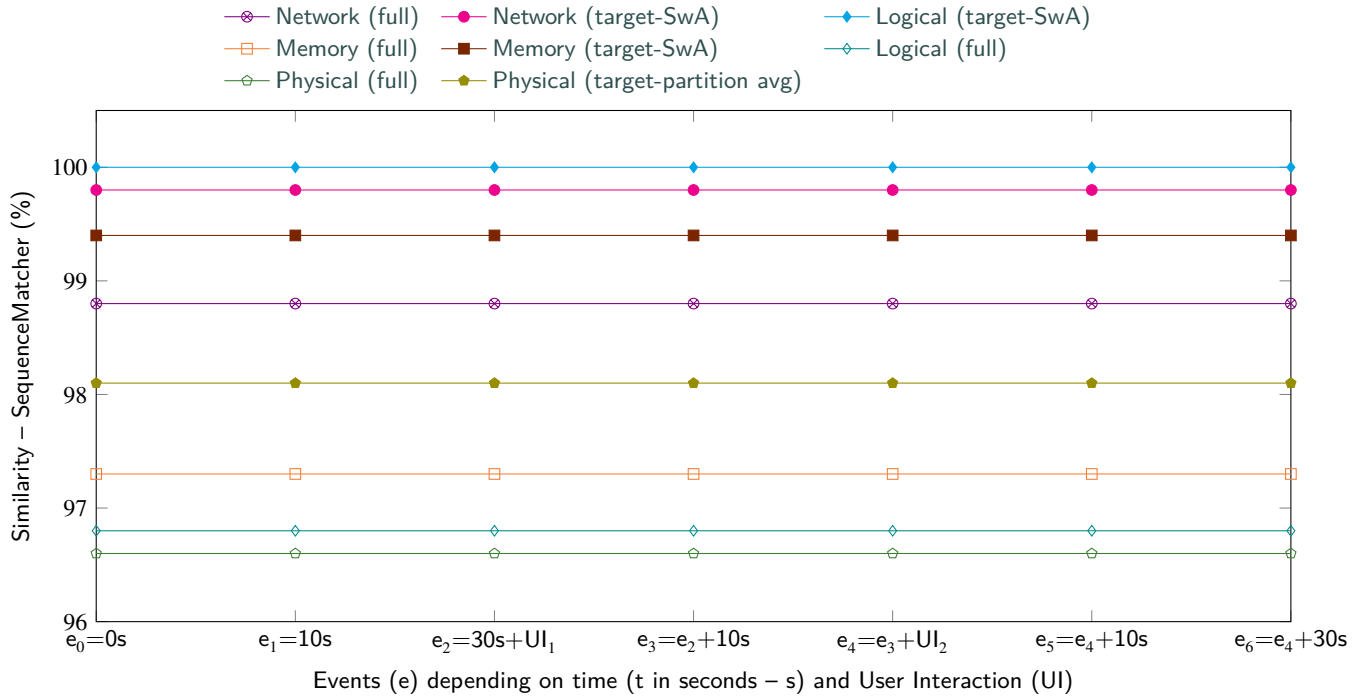
\begin{figure*}[pos=t]
\centering

\ref{acq-legend}\vspace{0.5ex}

\begin{tikzpicture}
\begin{axis}[
  width=\linewidth,
  height=0.48\linewidth,
  xlabel={Events (e) depending on time (t in seconds -- s) and User Interaction (UI)},
  ylabel={Similarity -- SequenceMatcher (\%)},
  xmin=0, xmax=6,
  ymin=96, ymax=100.5,
  xtick={0,1,2,3,4,5,6},
  xticklabels={
    e$_{0}$=0s,
    e$_{1}$=10s,
    e$_{2}$=30s+UI$_{1}$,
    e$_{3}$=e$_{2}$+10s,
    e$_{4}$=e$_{3}$+UI$_{2}$,
    e$_{5}$=e$_{4}$+10s,
    e$_{6}$=e$_{4}$+30s
  },
  xticklabel style={align=center},
  legend to name=acq-legend,                    
  legend cell align=left,
  legend columns=3,
  legend style={
    /tikz/every even column/.append style={column sep=6pt},
    draw=none
  },
  clip=false
]

\addplot[color=violet, mark=otimes] coordinates
  {(0,98.8) (1,98.8) (2,98.8) (3,98.8) (4,98.8) (5,98.8) (6,98.8)};
\addlegendentry{Network (full)}

\addplot[color=magenta, mark=otimes*] coordinates
  {(0,99.8) (1,99.8) (2,99.8) (3,99.8) (4,99.8) (5,99.8) (6,99.8)};
\addlegendentry{Network (target-SwA)}

\addplot[color=Cerulean, mark=diamond*] coordinates
  {(0,100) (1,100) (2,100) (3,100) (4,100) (5,100) (6,100)};
\addlegendentry{Logical (target-SwA)}

\addplot[color=Orange, mark=square] coordinates
  {(0,97.3) (1,97.3) (2,97.3) (3,97.3) (4,97.3) (5,97.3) (6,97.3)};
\addlegendentry{Memory (full)}

\addplot[color=Brown, mark=square*] coordinates
  {(0,99.4) (1,99.4) (2,99.4) (3,99.4) (4,99.4) (5,99.4) (6,99.4)};
\addlegendentry{Memory (target-SwA)}

\addplot[color=Cyan4, mark=diamond] coordinates
  {(0,96.8) (1,96.8) (2,96.8) (3,96.8) (4,96.8) (5,96.8) (6,96.8)};
\addlegendentry{Logical (full)}

\addplot[color=OliveGreen, mark=pentagon] coordinates
  {(0,96.6) (1,96.6) (2,96.6) (3,96.6) (4,96.6) (5,96.6) (6,96.6)};
\addlegendentry{Physical (full)}

\addplot[color=olive, mark=pentagon*] coordinates
  {(0,98.1) (1,98.1) (2,98.1) (3,98.1) (4,98.1) (5,98.1) (6,98.1)};
\addlegendentry{Physical (target-partition avg)}

\end{axis}
\end{tikzpicture}

\caption{Acquisition percentages over time (t) and user interaction (UI). The graph shows, for the eight different forensics acquisition methodologies, the similarities between the acquired analyzed data between the mDT and the real device under the same running condition (same architecture, OS, running application and version, running time, and user inputs). Similarity has been computed with the \textit{Ratcliff/Obershelp} algorithm}
\label{fig:acquisition_results}
\end{figure*}

\textit{\textbf{Network Acquisition and Analysis}}. Monitoring the network traffic by recording and saving in the \textit{pcap} file, the exchanged packets at every protocol level and using different protocols. For both systems (the real and the mDT), the tool {\tt tcpdump} is used for the acquisition, and tools, \eg { \tt Wireshark}, {\tt tshark}, {\tt Zeek}, Python libraries \eg { \tt Scapy} can be used for the analysis of the saved pcap file. Target network analysis on the SwA can be done with tools like {\tt pcapdroid}\footnote{\url{https://github.com/emanuele-f/PCAPdroid}} but also with {\tt tcpdump} selecting the UID of the target application found with {\tt iptables}, saving the content in a pcap file analyzed as the full network dumps.

\textit{\textbf{Memory Acquisition and Analysis}}. The RAM analysis is the most challenging one because the volatile nature of the RAM yelds its content highly unstable, changing over time as the system loads the content of a variable in memory or frees it only when needed, thanks to the use of the garbage collector. 
The memory acquisition can be made \emph{(i)} at the level of a target application, \ie extracting only the content of the RAM allocated to that specific application; and \emph{(ii)} dumping the complete RAM to monitor the whole system in the RAM and track the effects on the system from the main memory after a specific behavior. The first one can be done with tools like {\tt Fridump}\footnote{\url{https://github.com/Nightbringer21/fridump}} based on the instrumentation of the app via {\tt Frida}, analyzed with strings or simple hexadecimal view or specific encodings (\,  e.g., strings, images, audio, etc.). On the other hand, the complete acquisition can be made with {\tt LiME}\footnote{\url{https://github.com/504ensicsLabs/LiME}}. In this case, an extra step is needed, with the correct compilation of the kernel, to get a map of the memory and understand how the OS manages the memory and its content. The kernel can be compiled both in the emulated device and in the real device following the same procedure. The extracted content of the RAM called \textit{image of the RAM} can be analyzed with the tool {\tt Volatility}\footnote{\url{https://github.com/volatilityfoundation/volatility}}. 

\textit{\textbf{Logical Acquisition and Analysis}}. Extracting all files and directories accessible by the user (\eg those belonging to each user, application, or system) that is done as a copy of the non-deleted content.  
This can be done only for a specific target application or for the full file system. The acquisition is based on the mere copy of the directory under root permissions in a directory without root (\eg sdcard) so that the content can be downloaded via {\tt adb pull} command (accessible only for directories without the super-user access). At the same time, files and directories not protected by super-user privileges can be dumped on a local host without any transfer on the non-root directory (\eg sdcard). The analysis can be done for both systems with text editor tools or by analyzing the hexadecimal structure and binary similarities techniques.

\textit{\textbf{Physical Acquisition and Analysis}}. A complete bit-by-bit copy of the evidence under analysis to retrieve even the deleted content not overwritten in memory or retrieve its footprint from the filesystem. Forensically, this is the best acquisition methodology because, due to the perfect bit-by-bit copy, the analysis would be the same as done in the evidence but could be repeatable (one of the forensics principles). In a real device, the bit-by-bit acquisition can be done with specialized commercial tools such as {\tt Ufed}, which cannot be used in an emulator due to the specifics of {\tt UFED}. A free alternative in the real device, that does not required a paid license,  is the acquisition of the \textit{mmcblk0} using the \textit{dd} command {\tt adb exec-out su -c "dd if=/dev/block/mmcblk0" > memdump.img}. However, this partition cannot be found on the emulated device. The partition of the mDT strictly depends on the emulator tool (\eg Android Studio\footnote{\url{https://developer.android.com/studio}} and Genymotion\footnote{\url{https://www.genymotion.com/}}) and on the selected device and OS version. In the emulated device, the physical acquisition can be done with the same {\tt dd} command but considering the \textit{loop} and \textit{sda} partitions in the \textit{/dev/block} path. Each of them contains data about specific functionalities, such as application data, system data, etc. For this reason, to have the complete dump, all of them must be copied; otherwise, the analysis must be conducted only on the interesting target SwA.

\section{Experimental results and discussion}
\label{sec:discussion}
This section shows the results of the forensics acquisition methodology described in Section~\ref{sec:mobileliveforen:androdt}, highlighting what is the difference in the analysis between the dumped content from a real Android-rooted device and its mDT. Tests were done using Android Studio emulated with an ARM $64$ chipset as the real device (using a Macbook M3 computer). The real device and the mDT run with the same architecture (\ie arm$64$v$8$a) and with the same OS and SDK version (\ie Android $14$, API $34$). To demonstrate the consistency of the results, the experiments were repeated $10$ times, following the same conditions to prove the stability of the system. The same SwA with the same version has been executed simultaneously on both devices, with the same login and same interactions to guarantee data synchronization. Simultaneous interaction is possible with {\tt adb} when specifying the device with the option \textit{-s DEV\_NAME}. Data has been sent and intercepted with {\tt Frida}.

Figure~\ref{fig:acquisition_results} shows that data acquired from the mDT and the real device are quite similar. The content is unchanged between them, but sometimes we could not have the $100\%$ similarity or the same hash because of the emulator, as described in Section~\ref{sec:dt} and Section~\ref{sec:mobileliveforen:androdt}. Some vendor devices are not emulated by tools like Android Studio and Genymotion, \eg specific internals of \textit{Samsung}. Notably, such data are emulated by other mechanisms in the mDT. For example, in a real Samsung device the contacts are managed by Samsung itself (as shown in Listing~\ref{lst:samsung-contacts}). In contrast, in an emulated device, they are managed by the emulator as shown for Genymotion emulating Samsung Galaxy S4 in Listing~\ref{lst:genymotion-contacts} and Android Studio emulating Google Pixel 9 in Listing~\ref{lst:emulator-contacts}. Some other apps developed by the vendors can be downloaded and installed accordingly in the mDT and subsequently accessed with the same login data for synchronization. At the same time, the emulator contains proprietary data for the mDT belonging to Genymotion or Android Studio, according to the used one. For example, such directories contain data on the communication between the emulator and the Genymotion control system (\ie com.genymotion.genyd) or preferences and configuration options set by the user or system (\ie com.genymotion.settings). Data that cannot be emulated do not contain interesting information for the forensics investigation but just system and proprietary functions and internals to manage the pure emulation process.

\begin{lstlisting}[language=bash, basicstyle=\scriptsize, 
caption={Listing of contact-related packages in /data/data directory for Samsung Galaxy S4 real device},
label={lst:samsung-contacts}]
a33x:/data/data # ls | grep "contacts"
com.samsung.android.app.contacts
com.samsung.android.providers.contacts
com.sec.android.widgetapp.easymodecontactswidget
\end{lstlisting}

\begin{lstlisting}[language=bash, basicstyle=\scriptsize, 
caption={Listing of contact-related packages in /data/data directory for emulated Samsung Galaxy S4 with Genymotion},
label={lst:genymotion-contacts}]
a33x:/data/data # ls | grep "contacts"
com.android.contacts
com.android.providers.contacts
\end{lstlisting}

\begin{lstlisting}[language=bash, basicstyle=\scriptsize, 
caption={Listing of contact-related packages in /data/data directory for Pixel 9 API 35 (emu64a)},
label={lst:emulator-contacts}]
emu64a:/data/data # ls | grep "contacts"
com.android.providers.contacts
com.android.providers.contacts.
auto_generated_rro_product__
com.google.android.contacts
\end{lstlisting}

For this reason, the emulators do not manage SwA and services developed by the vendor that are always found in a real device. On the other hand, an emulator has additional services and SwA managing the emulator itself (\ie directories with the name \textit{androidstudio} or \textit{genymotion} according to the used emulator). Such directories cannot be deleted because they are essential for the mDT to work and emulate Android internals. Despite this, they do not affect the acquisition of forensics in terms of data integrity and data content (\ie user data and effects of the apps in the whole system). Other differences rely on the memory offset and network connection management while the data content is exactly the same. These are not limitations, as the emulated device contains the same data as the real device when a specific amount of time or event happens (as illustrated in Figure~\ref{fig:acquisition_results}). At the same time, the management of the emulation process does not affect these actions, and the related directories do not contain data that is useful for forensic investigation. 

The {\tt SHA$256$} hashing algorithm is commonly used in DF to ensure file integrity, as it provides collision-resistant identification of data. Ideally, hashes verify that forensic acquisitions from mDTs precisely match real devices. However, in practice, identical hashes are rarely achieved due to subtle differences in directory structures, hardware architectures (ARM vs. Intel), timing discrepancies, and memory management variations between real Android devices and their emulated counterparts. Despite different hashes, mDT reliably replicate real devices' behavior, memory content, file structures, and API interactions. Traditional comparison methods like {\tt ssdeep} hashes and Linux's {\tt diff} are ineffective because of architecture-induced offsets. Instead, the Ratcliff/Obershelp\footnote{\url{https://xlinux.nist.gov/dads/HTML/ratcliffObershelp.html}} algorithm implemented in Python's \textit{SequenceMatcher} of {\tt difflib}\footnote{\url{https://docs.python.org/3/library/difflib.html\#module-difflib}} is recommended for assessing content similarity. This method calculates a similarity ratio based on matching characters, typically yielding over $95$\% similarity, effectively confirming content equivalence despite minor structural or temporal discrepancies. The Ratcliff/Obershelp similarity ratio $R$ between two sequences $A$ and $B$ is defined as:
\begin{equation}
R = \frac{2M}{|A| + |B|}
\end{equation}
where $M$ is the total number of matching characters in the longest common subsequences found recursively, and $|A|$, $|B|$ are the lengths of the two sequences. This ratio effectively measures binary similarity, especially when structural variations exist but content remains functionally equivalent.

Notably, it is fundamental to highlight the differences between the emulation under Intel x86\_64 chipset or when using ARM. From the target SwA point of view there is no difference in terms of found data in the acquired evidence. This is because the app behavior is the same, except for the use on native libraries (\ie in ARM or Intel according to the relative architecture). From the full system point of view, the functioning mechanism is exactly the same, except for the different internals and system libraries, according to the target emulator. Hence, there is a difference in terms of hash or Ratcliff/Obershelp as just explained. 

The forensics analysis on the mDT allows to monitor and reconstruct the full activity of the user, the effects of an application in the system, considering a single target SwA or the full OS (\ie approach oriented to preserve the user privacy or a threat detection approach in the full system).

\textit{\textbf{Network Analysis}}. In this case, the hash analysis is different, while the content of the packets is exactly the same. Using the \textit{Ratcliff/Obershelp} algorithm, the network captures are almost $100$\% equals. Some packets differ because of the different internals of the emulator to manage some connections (only regarding the number of flows, not its content, nor the sent/received payload). The latter includes, for example, network discovery requests, handshake packets, or Googlecast requests in the case of the mDT. The same results were obtained for the single target application network traffic analysis. This means that it is possible to reconstruct the user's network connection in any way with a target analysis of the interesting application, hence preserving its privacy. Conversely, a complete network analysis of the system traffic can be made. This analysis consents to know, for example, the visited websites, the timeframe, and the sent/received files (with also the possibility of extracting them).

\textit{\textbf{Memory Analysis}}. About the target SwA analysis, the hash is different because of the different offsets and acquisition time. Still, the content is exactly the same as computed with the \textit{Ratcliff/Obershelp} algorithm, and the same data content, such as target strings of specific behaviors, can be identified.
In the complete RAM analysis, the hash is different. But the content of data (\eg files, processes, variables, etc.) extracted with Volatility is exactly the same. Therefore, the hash of the extracted files matches $100$\%. 
With this analysis, it is possible to dump the list of the running applications at a specific moment, how the OS manages them, and extract data from them (\eg encryption keys, one-shot pictures sent in chats, hidden behaviors). 

\textit{\textbf{Logical Analysis}}. For the target SwA, the content of its directory is completely the same between the mDT and the real device. In fact, the hash is the same, and the Ratcliff/Obershelp algorithm gives $100$\% of similarity. The same can be found for the other directories managed by the SwA, such as those in the \textit{sdcard} to save sent or received multimedia. For the full logical acquisition, because of the specific directories of the real device managed by the vendor or because of the specifics of the emulator, the hash is different, while the \textit{Ratcliff/Obershelp} highlights a very high similarity (about $97$\%).
The logical analysis allows to monitor the creation and modification of specific files, both at the target SwA level or the whole filesystem.

\textit{\textbf{Physical Analysis}}. In this case, due to the different files needed by the emulator and the real device, the hashes are different, but the content of the target behavior or applications is exactly the same. This analysis is very important as it allows to keep track, retrieve and reconstruct deleted files.

\section{Android Chat Case Study}
\label{sec:chat}
\begin{figure}[pos=t]
  \centering
  \begin{subfigure}[t]{0.48\linewidth}
    \centering
    \includegraphics[width=\linewidth,height=0.27\textheight,keepaspectratio]{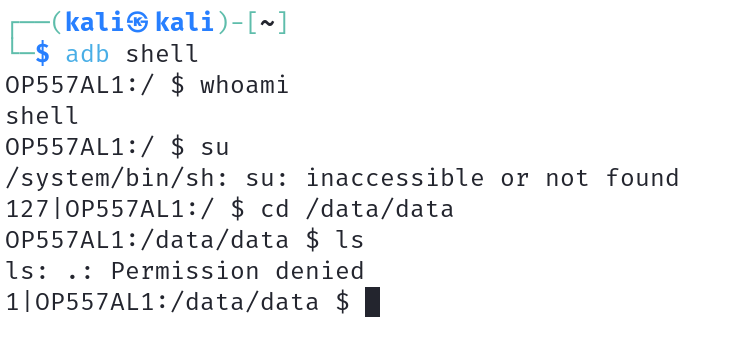}
    \caption{Access to \texttt{/data/data} on a non-rooted device}
    \label{sec:dt:fig:nonroot}
  \end{subfigure}\hfill
  \begin{subfigure}[t]{0.48\linewidth}
    \centering
    \includegraphics[width=\linewidth,height=0.27\textheight,keepaspectratio]{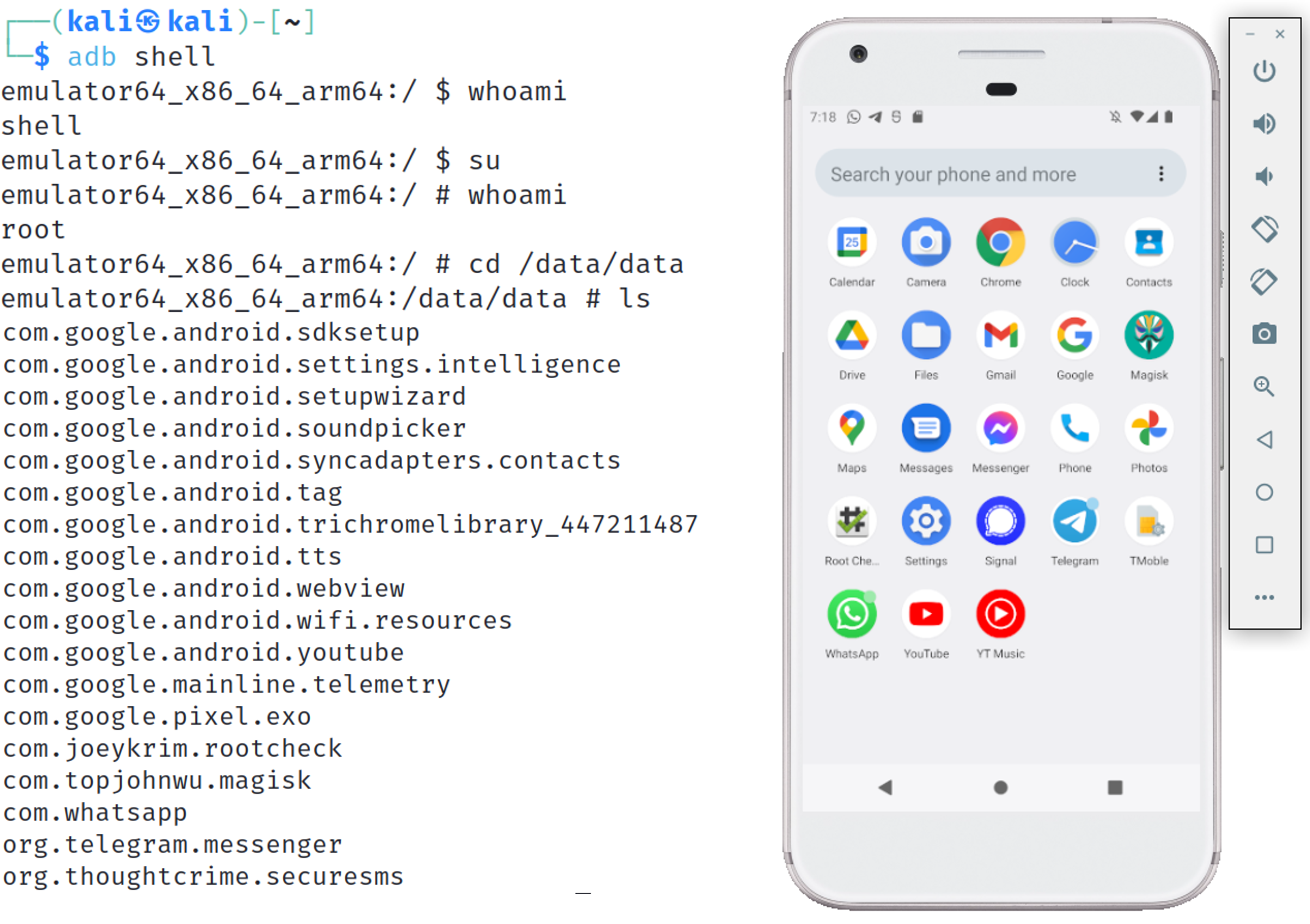}
    \caption{Access to \texttt{/data/data} on a rooted mDT}
    \label{sec:dt:fig:root}
  \end{subfigure}
  \caption{Protected directories on non-rooted device (above/up) vs. rooted Digital Twin (below/down)}
  \label{sec:dt:fig:dt}
\end{figure}

In this experimental case study, we installed the most downloaded SwA for instant messaging, \ie \texttt{WhatsApp, Telegram, Messenger}, in the rooted mDT. After login with an existent user, we checked the content of the super-user directories, \eg {\tt /data/data} directory. This experiment can be done in a real case scenario only if the defendant collaborates. We must unlock the real device and allow the sign-in procedure to be done on another device. After the installation and login, we analysed the content of the directory {\tt /data/data}. We used the real personal data of this article's first author and for privacy will not be displayed. It contains all data of the installed apps, and the content of the messaging SwA, as shown in Table~\ref{sec:dt:fig:root}. The SwA were accessed from a rooted real Android device, a Samsung A$33$, using the same credentials used in the physical device. The content was identical to the one in the mDT. Moreover, we successfully ran the command to dump a copy of the block devices with {\tt adb exec-out su -c "dd if=/dev/block/vdX" > vdX.img} (where X states for the different block devices) as done for the physical real device. These copies can be successfully analysed with programs like {\tt Autopsy}, which can obtain data from databases, saved pictures, etc. Alternatively, this analysis can be done with an {\tt adb} root shell ({\tt adb shell; su; cd data/data/appDirectory} and navigate to the directories as usual in Linux devices) by inspecting the directories and pulling the files with {\tt adb pull /data/data/APKDIR/FILE LOCAL\_PTH}. Saved multimedia files can be found only if the user saves them or the automatic saving is enabled. In particular, Telegram and Messenger allow multimedia file saving even before the mDT creation and app login. Conversely, WhatsApp allows content saving only for multimedia sent after the mDT creation and access. At the same time, the WhatsApp messages can be read even prior to the mDT creation. 

With a RAM forensics analysis it is possible to retrieve old and deleted messages and one-shot pictures (\ie images that can be visualised only for a specific amount of time chosen by the sender or just one). This can be done even in secret chats, both about sent and deleted messages or already visualised messages and even sent and visualised pictures or already deleted (\ie when visualisation time finished). Everything is found in clear text because the encryption key is saved in cache.

\begin{figure}[pos=t]
    \centering
    \includegraphics[width=0.6\linewidth]{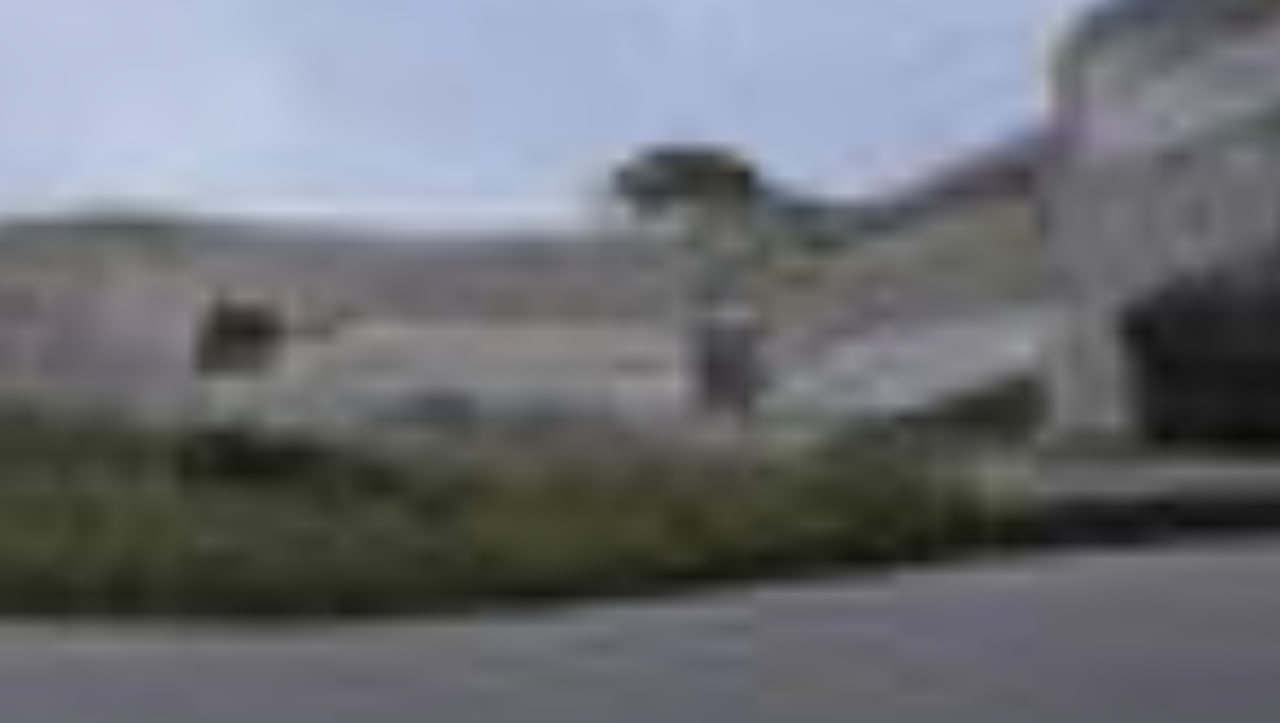}
    \caption{Image retrieved with carving methodology from the memory dump of the Telegram application in the mDT but the image visualized in the real device. The context of the image can be clearly and unequivocally seen}
    \label{fig:1shotretrieved}
\end{figure}

Detailed tests have been conducted on Telegram application on the chat's of the first author that for privacy reasons cannot be displayed. Pictures with one-shot visualization can be extracted and analyzed with a forensics analysis on the RAM. Even if the encryption key is stored in the cache directory of the mDT, the visualized one-shot image is in clear on the RAM. We made experiments also visualizing the image in the real device but dumping the RAM from the mDT. For this reason, by dumping the RAM, the one-shot visualization image can be retrieved using extraction tools such as {\tt binwalk} or \texttt{foremost}. Even if the quality was not $100$\% clear (\eg the image in Figure~\ref{fig:1shotretrieved} where we can clearly understand the context, in this case a landscape but in case of a nude it can be clearly detected), we managed to retrieve the one-shot picture visualized in the real device but extracted from the memory dump of the mDT. Due to the restriction of specific messaging SwA, some chats (\eg secret chats for Telegram) or the whole conversation can be visualized only on the device that created it. Here, the use of mDT is not helpful and a more specific methodology not discussed in this paper must be implemented. However, if sending a one-shot picture, thanks to the synchronization of the mDT and the real device, by dumping the mDT's RAM, the image can be retrieved even if visualized in the real device. 

The mDT can be used also to acquire the real device in the moment after the incident (\ie personal scenarios) even if not rooted. The rooted emulator can be used as a new device to transfer the data from the unrooted real device as if done with the backup and synchronization of a new real device. To do so, we need the Google Backup active, hence the collaboration of the owner. On the mDT, we need the app to allow the synchronization and access using the same Google account. In this experiment, we used a Samsung A21S non-rooted and its mDT emulator-5554. By doing so, as the mDT is rooted, we can directly access all data stored on the physical device but from the root emulator. To prove this, we used a rooted real device (Samsung A33) and compared the files of the messaging target applications (emulator-5556). We could see that the hashes of the files (\ie from the real device Samsung A33, its rooted mDT emulator-5556 and the mDT of a physical non-rooted device Samsung A21S whose mDT is emulator-5554) were exactly the same, except for few databases files where the device model is involved. In fact, for these files the Ratcliff/Obershelp coefficient is 98\% and Figure 6 shows the match of the first bytes. 
We can see that the acquisition produces the same output. We demonstrate that acquiring from a rooted mDT is equivalent in terms of content and data to acquire the real rooted device. 
Hence, our approach can be used also in this scenario.

\begin{figure}[pos=t]
  \centering
  \begingroup
  \scriptsize                      
  \lstset{basicstyle=\ttfamily\scriptsize, 
         columns=fullflexible,
         breaklines=true,
         keepspaces=true,
         showstringspaces=false,
         frame=single}

  \begin{subfigure}[t]{0.48\textwidth}
    \caption{Hash check (\texttt{sha1sum}) of the Telegram databases on:
    real rooted device (Samsung A33), its mDT (emulator--5556), and a rooted mDT with
    synchronized Telegram access (emulator--5554).}
    \vspace{2pt}
\begin{lstlisting}
user@linux DT % sha1sum DT_a21_root/telegram_5554/cache4.db
ba65af36973bb4cb831868ec4882ce204bfbf597
user@linux DT % sha1sum A33_root/cache4.db
ba65af36973bb4cb831868ec4882ce204bfbf597
user@linux DT % sha1sum DT_a33/telegram_5556/cache4.db
ba65af36973bb4cb831868ec4882ce204bfbf597
\end{lstlisting}
  \end{subfigure}\hfill

  \begin{subfigure}[t]{0.48\textwidth}
    \caption{Hexdump of the same Telegram databases (tail).}
    \vspace{2pt}
\begin{lstlisting}
user@linux DT % hexdump DT_a21_root/telegram_5554/cache4.db | tail
0000000 5153 694c 6574 6620 726f 616d 2074 0033
0000010 0010 0202 4000 2020 0100 0000 0100 0000
*
0000060 2e00 1b5f 00d0 0000 1000 0000 0000 0000
0000070 0000 0000 0000 0000 0000 0000 0000 0000

user@linux DT % hexdump A33_root/cache4.db | tail
0000000 5153 694c 6574 6620 726f 616d 2074 0033
0000010 0010 0202 4000 2020 0100 0000 0100 0000
*
0000060 2e00 1b5f 00d0 0000 1000 0000 0000 0000
0000070 0000 0000 0000 0000 0000 0000 0000 0000

user@linux DT % hexdump DT_a33/telegram_5556/cache4.db | tail
0000000 5153 694c 6574 6620 726f 616d 2074 0033
0000010 0010 0202 4000 2020 0100 0000 0100 0000
*
0000060 2e00 1b5f 00d0 0000 1000 0000 0000 0000
0000070 0000 0000 0000 0000 0000 0000 0000 0000
\end{lstlisting}
  \end{subfigure}

  \vspace{3pt}
  \caption{Telegram \texttt{cache4.db} equality across devices: identical SHA-1s (a) and matching hexdump tails (b) for the real rooted device (Samsung A33), its synchronized mDT (emulator--5556), and a rooted mDT with synchronized Telegram access (emulator--5554).}
  \label{fig:telegram-db}

  \endgroup
\end{figure}

In case the defendant does not collaborate and we cannot get access on the real device because is not rooted, we can install the same application in the rooted mDT and manually try to understand where the app stores data and be sure that the interesting results can be recovered with root.

During our tests, the methodology was successfully applied to several apps. However, it must be noted that it does not work for applications that rely exclusively on a local database for account monitoring, such as Signal: when accessing the same Signal account from another device, no messages are imported, which limits the effectiveness of the approach for this specific case.

\section{Research Challenges and Open Issues}
\label{sec:challenges}
The proposed MoLIFE methodology strictly depends on the technology used. For this reason, the current version of MoLIFE has some limitations related to its progress. 
Additionally, employing the MoLIFE system strictly depends on the country's regulations. In the case of Europe, for example, it must be GDPR compliant to respect privacy and AI Act. Moreover, international law enforcement collaboration must be regulated, and the legislation of each country involved in the investigation must be respected. To be fully integrated into our daily lives, some standards on technology must be followed. At the time of writing this paper, some studies have been carried out on standardization, as highlighted by \cite{alcaraz_digital_2025}.
The MoLIFE system has some intrinsic limitations that do not allow use in every scenario. As an example, the system cannot monitor applications that detect the execution on an emulated environment, changing its behavior accordingly (\ie MATE scenario). For example, some SwA cannot be monitored because of development restrictions that prevent it from being installed on devices with super-user access. It is the case of PayPal that cannot be installed from the Play Store, neither by {\tt adb} if the device is rooted. Accordingly, some modifications to the SwA via reverse engineering techniques can be made to bypass such checks. Moreover, self-developed SwA by the company or other entities to be used for a specific use case is blocked during installation by the preventive stage on the hash matching in the database. This problem can be overcome by modifying the database, \ie adding the hash of the SwA developed for the specific use case or adding the SwA in the official Store so that it is verified. This last option, sometimes for privacy restrictions, industrial secrets, or intellectual properties, cannot be adopted. Hence, modifying the database accordingly is the best solution. As highlighted in Section~\ref{sec:mobileliveforen:androdt} for the Android case study, some forensics methodologies and tools cannot be applied to iOS devices as they are less investigated due to the proprietary system's restrictions. However, the study by Hu \textit{et al.}\cite{Hu23_ACM} discovered that SwA behaviors and vulnerabilities in Android are found in the same way in iOS devices. Thus, some features and detection mechanisms could remain unchanged, but more detailed investigations are needed. Regarding the SwA monitoring, the MoLIFE system is designed to monitor only one SwA of interest. Monitoring more than one SwA leads to an additional complex layer because every change in each SwA must be tracked. Tests must also be conducted to consider more than one SwA. At the time of presenting the MoLIFE methodology, no dataset is available to test the system in a real-case scenario. Additionally, no robust selected features have been publicly released, nor have accurate models been used to detect the presented behaviors. Accordingly, more research is needed in this direction to extend the use of the proposed methodology to other types of devices such as mobile devices based on iOS. Moreover, at this stage of development of MoLIFE, it is difficult to test the robustness to adversarial attacks.  
The estimation of the robustness is essential for better accuracy of the system, and reduce the possibility that detection is eluded. For example, robustness is essential in some critical cases such as the AI-based detection of violence (or other sensitive topics) via the real-time of chats. Finally, the security and protection of the MoLIFE system, with respect to risks related to the infrastructure where the mobile devices are used, should be duly analysed and addressed. For example, the protection of data over the communication channels (\eg privacy preservation and data exfiltration, 5/6G communication security challenges), and the protection of the edge from attacks, and the protection of the mDT from unauthorized access and privilege escalation by the attacker, as highlighted in \cite{alcaraz_digital_2022}, should be better analyzed and addressed.

\section{Conclusion}
\label{sec:conclusion}
This paper proposes a new mobile device forensics methodology (MoLIFE) based on new technologies. MoLIFE will analyze extracted interesting data in case of suspicious activity such as a cyber attack or a traditional crime to the involved owner. The methodology aims at overcoming the limitations of modern mobile devices that cannot be fully analyzed if no super-user privilege has been acquired before the rising alarm of a running potential threat. Moreover, it is safe not to previously acquire the super-user privilege on the physical devices because this could expose them to more vulnerabilities and attacks. In this case, a mDT with admin privileges can help monitor and analyze. The monitoring stage can be controlled by AI algorithms, specifically trained to the presented analysis and robust to adversarial attacks, with the capability of explaining and reporting to the human forensics analyst the decision while keeping the real device as protected as possible. Moreover, this paper presents a case study on an Android system, showing how the forensic analysis of an mDT is the same as that of a real Android device. The proposed system has some limitations mainly related to the limitations of the current state of the technologies. The proposed system has the potentiality to evolve in agreement to the advance of the employed technologies

\section*{Acknowledgments}
This work was partially supported by Project SERICS (PE00000014) under the NRRP MUR program funded by the EU - NGEU. This work was carried out while Silvia Lucia Sanna was enrolled in the Italian National Doctorate on Artificial Intelligence run by Sapienza University of Rome in collaboration with the University of Cagliari. This work was designed with the collaboration of Davide Maiorca and Leonardo Regano.

\bibliographystyle{cas-model2-names}
\bibliography{\jobname}

\section*{Biographies}
\vspace{-2mm}
\bio{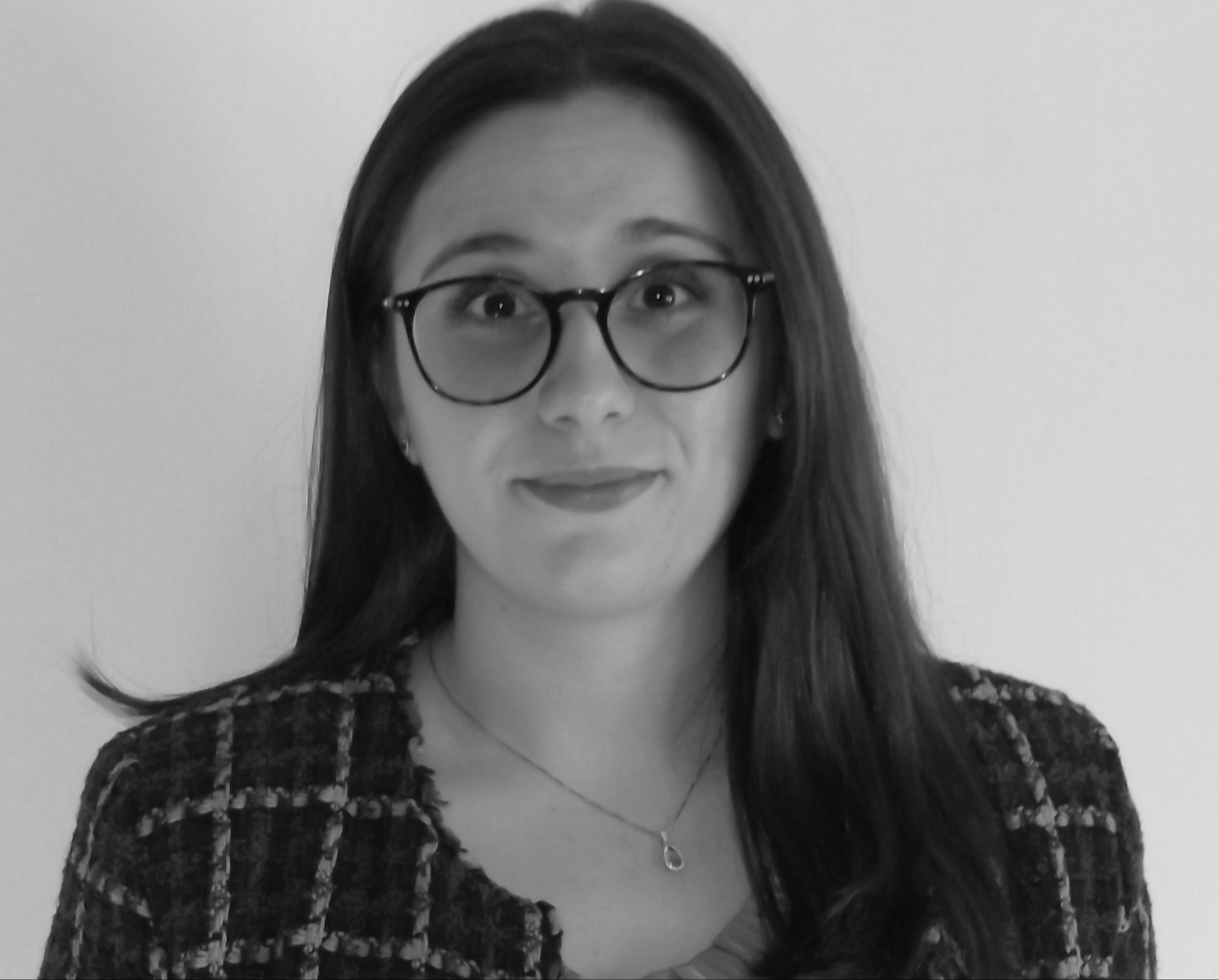}
\textbf{Silvia Lucia Sanna} is a Ph.D. student (since November 2022) in the National Ph.D. in AI for Security and Cybersecurity. Her research focuses on AI-driven vulnerability and threat detection on Android using digital forensics features. In 2023 she served as Youth Ambassador for Women4Cyber Italy.
\endbio
\vspace{-2mm}
\bio{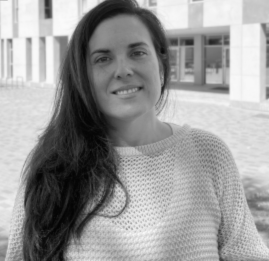}
\textbf{Cristina Alcaraz} is an Associate Professor in the Department of Computer Science at the University of Malaga. Her research spans the security of cyber–physical systems, IIoT, and digital twins, with applications to Industry 5.0, manufacturing, supply chains, and smart grids.
\endbio
\vspace{-2mm}
\bio{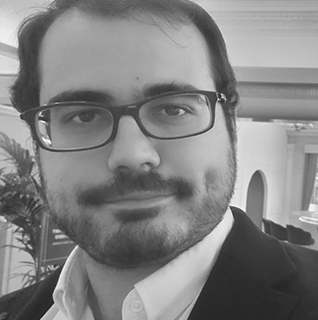}
\textbf{Alessandro Sanna} is a Postdoctoral Researcher at the University of Cagliari and a member of the Pattern Recognition and Applications Laboratory. His work focuses on malware detection and threat intelligence, with emphasis on Living-off-the-Land malware and ML/DL for threat detection.
\endbio

\bio{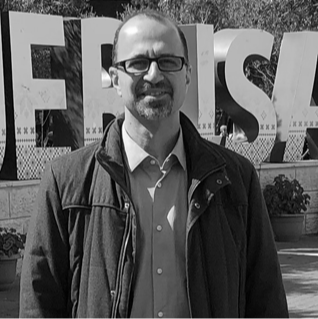}
\textbf{Giorgio Giacinto} is a Professor of Computer Engineering at the University of Cagliari and coordinates the M.Sc. program in Computer Engineering, Cybersecurity, and AI. His research centers on machine learning for malware detection (180+ publications). He is Editor-in-Chief of the \textit{Security Engineering and Applications} section of the \textit{Journal of Cybersecurity and Privacy}, holds leadership roles in Italy’s Cybersecurity National Lab (CINI) and represents it in the European Cybersecurity Organization. He is an IAPR Fellow and a Senior Member of IEEE and ACM.
\endbio
\vspace{-2mm}
\bio{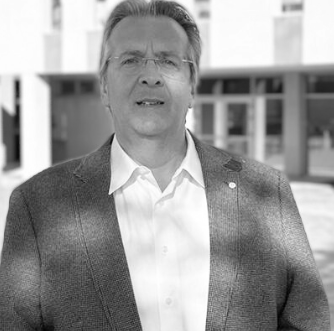}
\textbf{Javier Lopez} is a Full Professor in the Department of Computer Science at the University of Malaga. His research focuses on network security, security protocols, and critical information infrastructures, and he has led several national and EU research projects in these areas.
\endbio
\vspace{-2mm}

\end{document}